\documentclass[prx, aps, amssymb, twocolumn, superscriptaddress, notitlepage, longbibliography]{revtex4-2}

\usepackage{amsmath}
\usepackage{amsfonts}
\usepackage{graphicx} 
\usepackage{float}
\usepackage{braket}
\usepackage{xcolor}
\usepackage{subfigure}
\usepackage{cancel}
\usepackage{lineno}
\usepackage[colorlinks,linkcolor=blue,citecolor=blue,urlcolor=blue]{hyperref}

\begin{document}

\title{Quantum Chaos in Random Ising Networks}

\author{Andr\'as Grabarits\href{https://orcid.org/0000-0002-0633-7195}{\includegraphics[scale=0.05]{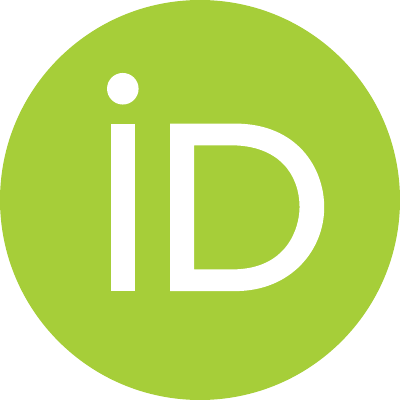}}}
\email{andras.grabarits@uni.lu}
\affiliation{Department  of  Physics  and  Materials  Science,  University  of  Luxembourg,  L-1511  Luxembourg, Luxembourg}

\author{Kasturi Ranjan Swain\href{https://orcid.org/my-orcid?orcid=0009-0008-7227-3856}{\includegraphics[scale=0.05]{orcidid.pdf}}}
\email{kasturi.swain@uni.lu}
\affiliation{Department  of  Physics  and  Materials  Science,  University  of  Luxembourg,  L-1511  Luxembourg,  Luxembourg}

\author{Mahsa Seyed Heydari}
\affiliation{Department of Physics, University of Konstanz, D-78457 Konstanz, Germany}

\author{Pranav Chandarana\href{https://orcid.org/0000-0002-3890-1862}{\includegraphics[scale=0.05]{orcidid.pdf}}}
\affiliation{Department of Physical Chemistry, University of the Basque Country UPV/EHU, Apartado 644, 48080 Bilbao, Spain}
\affiliation{EHU Quantum Center, University of the Basque Country UPV/EHU, Barrio Sarriena, s/n, 48940 Leioa, Biscay, Spain}

\author{Fernando J. G\'omez-Ruiz\href{https://orcid.org/0000-0002-1855-0671}{\includegraphics[scale=0.05]{orcidid.pdf}}}
\affiliation{Departamento de F\'isica Te\'orica, At\'omica y \'Optica and  Laboratory for Disruptive Interdisciplinary Science, Universidad de Valladolid, 47011 Valladolid, Spain}

\author{Adolfo del Campo\href{https://orcid.org/0000-0003-2219-2851}{\includegraphics[scale=0.05]{orcidid.pdf}}}
\affiliation{Department  of  Physics  and  Materials  Science,  University  of  Luxembourg,  L-1511  Luxembourg,  Luxembourg}
\affiliation{Donostia International Physics Center,  E-20018 San Sebasti\'an, Spain}

\begin{abstract}
We report a systematic investigation of universal quantum chaotic signatures in the transverse field Ising model on an Erd\H{o}s-R\'enyi network. This is achieved by studying local spectral measures such as the level spacing and the level velocity statistics. A spectral form factor analysis is also performed as a global measure, probing energy level correlations at arbitrary spectral distances. Our findings show that these measures capture the breakdown of chaotic behavior upon varying the connectivity and the strength of the transverse field in various regimes. We demonstrate that the level spacing distribution and the spectral form factor signal this breakdown for sparsely and densely connected networks. The velocity statistics capture the surviving chaotic signatures in the sparse limit. However, these integrable-like regimes extend over a vanishingly small segment in the full range of connectivity. Our work elucidates the emergence of quantum chaos in relation to network topology, establishing a bridge between many-body physics and network theory. 
\end{abstract}

\maketitle

\section{Introduction}\label{sec:Intro}
Investigating the properties of quantum Ising models is of utmost importance given their widespread applications across various domains, including quantum optimization~\cite{Lucas_NPHard_sumamry, AlbashRevModAQC_NP_hard, MezardPArisis_ClSG_Sumamry, HartMann_Opt_Problems_2005, Barahona_Complexity_NP_Hard_1982,garey1979computers}, quantum chemistry~\cite{Qchem}, condensed matter physics~\cite{BravoPrieto2020scalingof}, and high energy physics~\cite{di2023quantum}. A pivotal aspect governing the behavior of Ising models lies in the connectivity of their underlying graphs. While many instances involve simple graph structures such as lattices or regular graphs, certain scenarios feature more complex networks with intricate characteristics~\cite{dynamic_cmplx_network_2008, Collective_dynamics_Watts1998, Newman_book, Dorogovtsev_Phase_Rans_IsingNetwork}.

A complex network can be represented by a random graph $\mathcal{G}(V, E)$, described by a set of vertices $V$ and a set of edges $E$, sampled from a given probability distribution. The examination of complex networks is of immense significance, as they appear in diverse fields, including ecology, biology, population dynamics, neural networks, and the World Wide Web~\cite{Jeong_Barabasi_Metabolic_2000, montoya2000complexity_Ecological, Pietronero_Ecological_Nework_2003, Bassett_Neural_Networks_2007, BARABASI_WWW_1999}. Disorder in these graphs arises from the randomness of the connectivity and the weights assigned to each edge.  This randomness is essential to investigate the statistical properties of complex systems with exponentially growing degrees of freedom~\cite{Erdos_Renyi_Graph_1, Erdos_Renyi_Graph_2, WS_RandomNetwork, Barabasi_1999_RandomGraph, Abiuso_2024, bollobás1985RandomGraphs}.

Studying the connection between complex networks and Ising models has attracted outstanding attention over the past decades. 
For instance, graph-based combinatorial optimization problems can be mapped to classical Ising Hamiltonians, such that the ground state encodes the solution~\cite{Lucas_NPHard_sumamry}.
Optimization problems as a ground-state search have been investigated intensely in the context of
the Quantum Annealing Algorithm (QAA)~\cite{Fahri_QAO_Science, Santoro_QAO_Science, PYoung_200_ExpGap_2, Bapst_QM_Spin_Glass_Review, Troyer2014}. In QAA, quantum fluctuations induced by the transverse field replace thermal fluctuations of classical annealing algorithms~\cite{KirkPatrick_SA}. Strong fluctuations delocalize the eigenstates over the computational basis, improving the performance of QAA. 

Spectral regions corresponding to these eigenstates are associated with quantum chaos
~\cite{Apollfo2024SFFQMChaosproposal, Vidmar_ErgodBreakHeisenberg_2020, Rigol_IntBreak_g_c, Rigol_UniversalOnSetChaos_2021, Santos_DisorderedHeisenberg_QMChaos2008, Haake_QM_Signatures_2001}.
In quantum chaos, energy level correlations follow the predictions of random matrix theory (RMT)~\cite{ProzenPRXKickedTOPSFF, Dubertrand_2016_LevelStatMBChaos,Mehta_1991_Random_Matrices_Review, GUHR_RMT_Review_1997}.
The tools of equilibrium statistical mechanics are applicable in these regimes. The well-established ergodic hypothesis and the eigenstate thermalization hypothesis (ETH) are also satisfied ~\cite{Deutsch_QMStatMech_ETH, Rigol2008_ETH, Deutsch_2018_ETH_Review,  Alessio_2016_ETH_Review}. 
However, integrable parts of the Hamiltonian may break down quantum chaos and localize the eigenstates. 
In the disordered case, this gives rise to many-body localization (MBL)~\cite{Huse_SpacingRatio_MBL, Abanin_RevMod_MBL_ETH, BASKO2006_Weak_Interacting_Fermions_MBL} with additional conserved quantities and local constants of motions.
In contrast to chaotic spectra, equilibrium statistical mechanics loses its validity in the MBL phase. Due to this property, it is crucial to identify the signatures of quantum chaos and its breakdown in favor of MBL. This has been addressed in several works, including experiments with cold atoms, trapped ions, and superconducting qubits~\cite{ Smith2016_Experiment_Cold_ion_MBL, Chaudhury2009_KickedTopQMChaos_Experiment}. 
  
\begin{figure*}[t]
\centering
\includegraphics[width=1\linewidth]{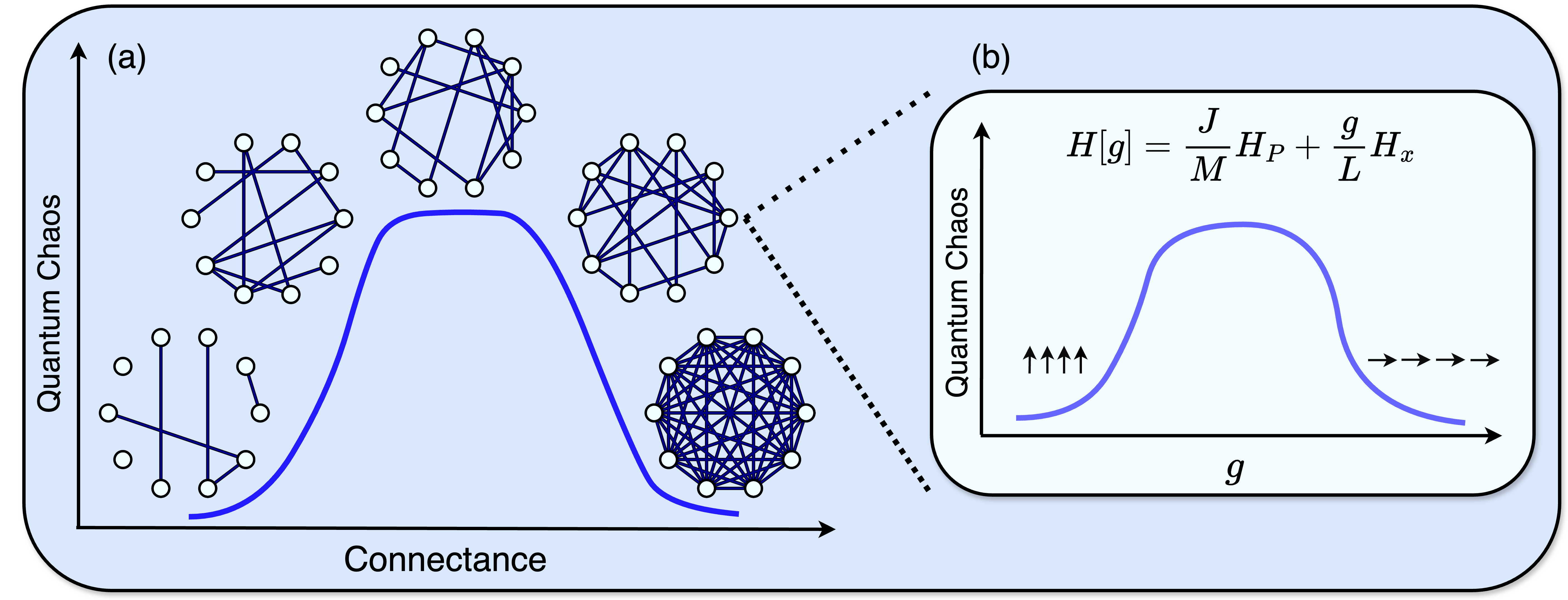}
\caption{\label{fig:ER} 
Schematic diagram showing an overview of the work. Each random graph $\mathcal{G}(L,M)$ corresponds to a transverse field Ising model with fixed connectance $\tilde M = \frac{2}{L(L-1)} M$.
$(a)$ Quantum chaos is explored as a function of the connectance $\tilde M$ ranging from the sparsely connected regime to the LMG limit. 
$(b)$ Signatures of quantum chaos are inspected by varying the transverse field and the connectance.}
\end{figure*}

However, establishing the boundary between quantum chaos and the MBL phase remains challenging. This comes with a numerical bottleneck as the thermodynamic limit is approached ~\cite{Panda_2019_MBL_Size, ABANIN2016SystemSize}. Comprehensive full-spectrum analysis has primarily been confined to Heisenberg models, with some exceptions extending to Ising spin chains~\cite{Pollmann_MBL, Laumann_QREM_MBL_1, Schardicchio_Bethe, ABANIN2016SystemSize,sherringtonKirkpatrickMBL, Pollmann_2021Fockaspaceview}.
Studies of spectral characteristics have mostly incorporated the disorder in the interaction. Nonetheless, the effect of a random graph topology has been restricted to the analysis of the spin glass phase~\cite{Bethe_Sping_Glass, Kim_2005_SGPhaseTrans_IsingNetwork, Herrero2009_SGPhaseTrans_IsingNetwork, shovan2024} and thermal phase transitions of classical Ising models~\cite{VianaLopes_SmallWorldPhaseTrans_Ising, CHatterje_PhaseTrans_IsingNetwork, Dorogovtsev_RevMod_PhaseTransIsingNetworks} near the ground state. There is a sizable literature on the eigenvalue statistics of the adjacency matrix itself~\cite{Alt2021_ERAdjMatrix_Deloc, Sade2005_LocERAdjacencyMatrix, Antti_AdjacencyLargeEigenvalue,cugliandolo2024multifractal_ER}. Furthermore, single-particle quantum chaos on random graphs has been studied both in one- and two-dimensional Anderson models~\cite{Garcia_Mata_PRL_RandomGraphAnderson, AndersonRandomRegular_1, AndersonRandomRegular_2} and in $2$-body Sachdev-Ye-Kitaev (SYK$_2$) models also touching its possible extensions to the many-body case~\cite{Dario_Impurity_SYK_Chaos, Dario_Operator_Growth_Chaos}.

In previous works, the disorder was mostly represented by continuous random variables rather than by discrete-valued randomness. However, the latter provides a more ideal testbed for experimental realizations~\cite{Discrete_Binary_Disorder_BoseHubbard, Discrete_Disorder_MBL, Discrete_Binary_Disorder, Monica_nature_2021, King2023, king2024computational}.
Additionally, plenty of questions remain unanswered. For instance, when does graph randomness lead to chaotic quantum behavior? Does the exponential suppression of the long-range delocalization exist beyond the short-range perturbative regime of the transverse fields? Is the onset of quantum chaos universal? 

In this work, we address these questions for an Ising model on a random graph network via the level spacing distribution, the level velocity statistics, and the spectral form factor. In particular, we identify the regimes of the graph connectivity, where many signatures of quantum chaos disappear. Universal conditions for the onset of quantum chaos in terms of the transverse field are also investigated via large-order perturbation theory. Quantum chaotic behavior is only absent in vanishingly small regions compared to the full range of the connectivities.
 
This work is organized as follows. In Sec.~\ref{Sec: On-site_Energies}, the structure of the spectral correlations in the computational basis is studied. In Sec.~\ref{sec:LStat}, the spectrum is investigated numerically by a nearest-neighbor level spacing analysis with analytical methods. In Sec.~\ref{sec:VelStat}, quantum chaos is probed via a further local measure, the level velocities. In Sec.~\ref{sec:SFF}, the spectrum is analyzed by the spectral form factor, diagnosing quantum chaos via long-range spectral properties. Finally, we devote Sec.\ref{sec:Concl} to conclusions and future directions.

\section{Preliminaries}\label{Sec: On-site_Energies}
This section provides details about the model and the underlying graph structure. Let us consider a general QAA Hamiltonian 
\begin{equation}
     H[g]=\frac{J}{M}\,H_P+\frac{g}{L}\,H_x\,, \label{eq2}
 \end{equation}
where, $H_x=-\sum_i\,X_i$, $H_P = -\sum_{i,j}\,A_{ij}Z_iZ_j$, and $g$ is the strength of the transverse field. Here, $A_{ij}$ is the adjacency matrix with $A_{ij}=1$ for connected vertices and zero otherwise. We consider the Erd\H{o}s-R\'enyi graphs $\mathcal G(L, M)$, where $(L,M)$ denotes the number of vertices and edges, as depicted in Fig.~\ref{fig:ER}. Every graph configuration is sampled with probability 
 $p_M=1/\begin{pmatrix}
 \binom{L}{2}\\
 M
 \end{pmatrix}$.
The prefactors $M$ and $L$ scale the bandwidth to the order of the energy unit $\mathcal O(J)$. 
The couplings are chosen to favor ferromagnetic ordering, $J>0$.  
To characterize the connectivity, we define the connectance $\tilde M =\frac{2}{L(L-1)} M$ \cite{Newman_book}, interpolating between the limits of disconnected graphs and the fully connected Lipkin-Meshkov-Glick model (LMG)~\cite{LMG1965}. The connectance varies in the interval $[0,1]$. 
 
We investigate the relation between the graph randomness and the statistics of the eigenenergies of $H_P$.
The Hamiltonian $H[g]$ can be realized as an ``Anderson-like" disordered hopping model on an $L$-dimensional hypercube~\cite{Altshuler_ExpGAp_3SAT, Schardicchio_Bethe},
\begin{equation}
    H[g]=\sum_k\,E_k\,\lvert k\rangle\langle k\rvert - \frac{g}{L} \sum_{\langle k,k^\prime\rangle}\lvert k\rangle\langle k^\prime\rvert\,.
\end{equation}
The disordered interactions generate the on-site energies $E_k$ with non-trivial correlations. The transverse field terms induce the nearest neighbor hoppings $\langle k,k^\prime\rangle$ via single spin flips along the edges.

In the case of random interactions, the average of the on-site energies can be shifted to zero, with the tunable energy variances characterizing the statistical properties.
However, with the more constrained graph randomness, the influence of the disorder cannot be described solely by the energy variances. These variances are bounded by the finiteness of $\tilde M$, and the statistical properties are strongly influenced by the non-trivial structure of the mean values.
Therefore, we characterize the conditions of the onset of quantum chaos by the generalized relative variances and correlations.
 As detailed in App.~\ref{App: Onsite_Energies}, these quantities take the form,
\begin{align}
    &\langle E_k\rangle_\mathrm{ER}\approx\tilde M\,\frac{S^2_k-L}{M}J\,,\\
    &\langle \delta E^2_k\rangle_\mathrm{ER}\approx2\tilde M\,(1-\tilde M)\,\frac{L^2}{M^2}J^2\,,\\
    &\langle\delta E_k\delta E_{k+r}\rangle_\mathrm{ER}\approx2\tilde M(1-\tilde M)\frac{L^2+6r(r-L)}{M^2}J^2\,.
\end{align}
 Here, $S_k$ denotes the total spin of the $k^{\text{th}}$ spin configuration, and $\langle\delta E_k\delta E_{k+r}\rangle_\mathrm{ER}$ is the correlator of states at Hamming distance $r$ averaged over the graphs.
Their most probable value varies as $S_k\sim\sqrt L$, yielding the typical scaling $\langle E_k\rangle_\mathrm{ER}\sim \tilde M\frac{L}{M}J$. In the thermodynamic limit, these relations become exact as the probability of a given link converges to $\tilde M$. The corresponding relative variances and correlations scale as
\begin{align}\label{eq:relative_correlators}
    &\frac{\sqrt{\langle \delta E^2_k\rangle_\mathrm{ER}}}{\langle E_k\rangle_\mathrm{ER}}\approx\sqrt{2\frac{1-\tilde M}{\tilde M}}\,,\\
    &\frac{\sqrt{\langle\delta E_k\delta E_{k+r}\rangle_\mathrm{ER}}}{\langle E_k\rangle_\mathrm{ER}}\approx\sqrt{2\frac{1-\tilde M}{\tilde M}}\sqrt{1+6\frac{r}{L}\left(\frac{r}{L}-1\right)}\,.
\end{align}
Here, the relative strength of the correlators can be characterized by a single average value between the configurations of $k$ and $k+r$ having the same leading order behavior in $L$.

\begin{figure*}[t!]
    \includegraphics[width=1\linewidth]{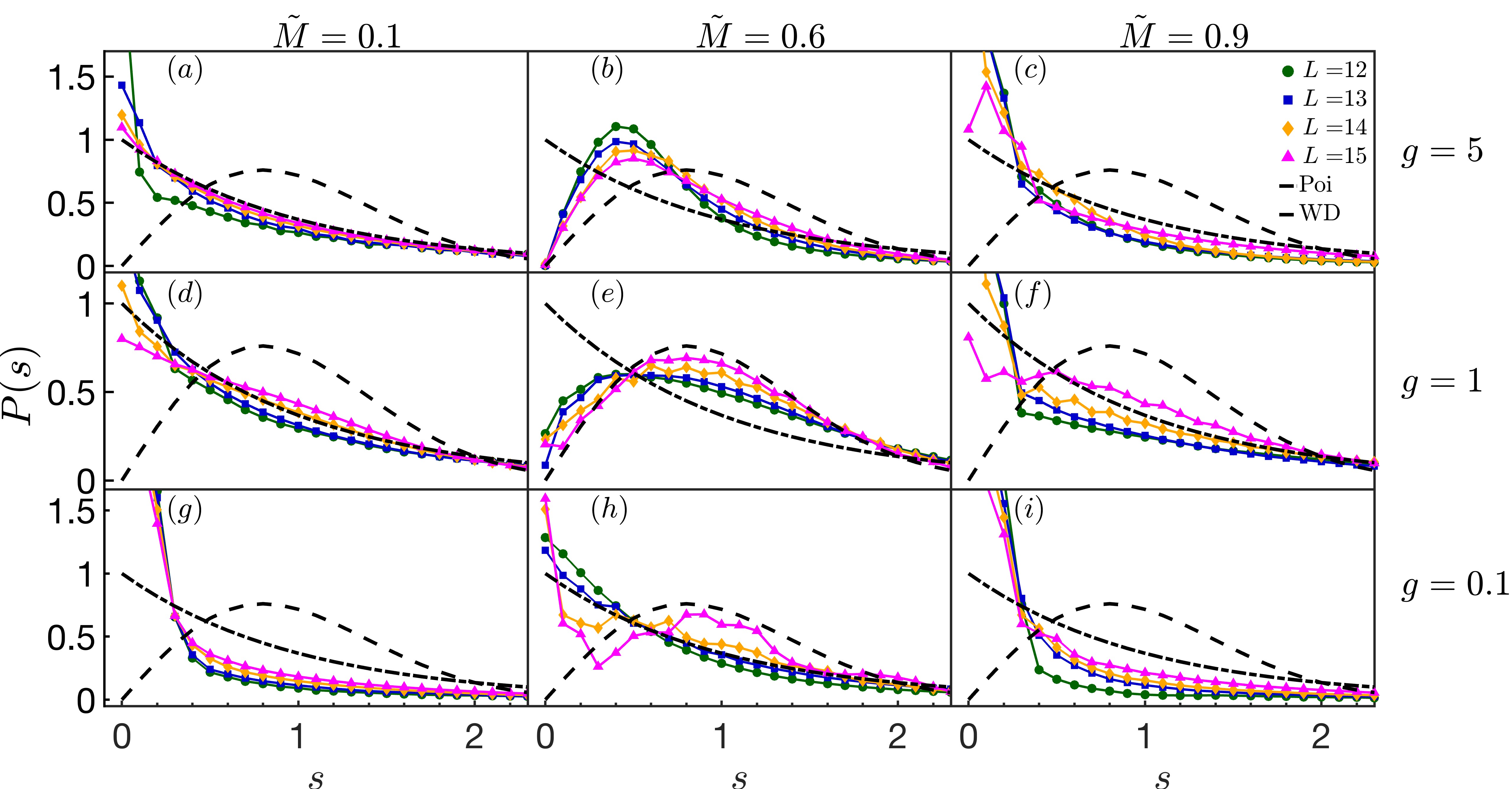}
    \caption{Nearest neighbor level spacing distribution in the perturbative $g=0.1J$, intermediate $g=J$, and strongly polarized $g=5J$ regimes, and for sparsely $\tilde M=0.1$, intermediate $\tilde M=0.6$, and densely connected $\tilde M=0.9$ graphs with $L=12,\,13,\,14,\,15$. The energy window $[0.3\,J,\,0.4\,J]$ was considered in the unfolded spectrum.
    Panels $(a)-(c)$: $g=5\, J$, displaying the perturbative breakdown of quantum chaos. Level repulsion gets stronger as $L$ is increased. This growth is slower for low and high $\tilde M$ in $(a)$ and $(c)$, respectively.
    Panels $(d)-(f)$: Similar behavior for $g=J$ with a faster restoration of quantum chaos for intermediate $\tilde M$.
    Panels $(g)-(i)$: $g=0.1\, J$, exhibiting similar features with level repulsion getting stronger as $L$ grows.
    }\label{fig: LStat_examples}
\end{figure*}

\section{Level spacing distribution }\label{sec:LStat}
In this section, local spectral analysis is performed via the statistics of the spacing between adjacent energy levels for various $g$ and $\tilde M$ values.

\subsection{Chaotic regime of connectance and transverse field}

The level spacing distribution, characterizing the difference between adjacent energy levels, constitutes a prominent tool in the analysis of spectral statistics \cite{Haake_QM_Signatures_2001}. It captures the chaotic and the localized regimes, as tested in various models~\cite{BASKO2006_Weak_Interacting_Fermions_MBL, Huse_SpacingRatio_MBL, Pollmann_MBL, Laumann_QREM_MBL_1}. 
We employ level spacing distribution at intermediate $\tilde M$, sufficiently outside the densely $1-\tilde M\sim1/L$ and the disconnected limits $\tilde M\sim 1/L$.

Quantum chaos is signaled when the level spacing distribution converges towards the Wigner-Dyson (WD) distribution. However, it breaks down when the level spacings are more likely to concentrate around zero. An example of this is the Poissonian statistics for uncorrelated energy levels. The corresponding distribution functions are

\begin{equation}    
        P_\mathrm{Poi}(s)\propto e^{-s},\quad P_\mathrm{WD}(s)\propto s\,e^{-\frac{\pi}{4}s^2},    
\end{equation}
with the level spacing scaled to have unit average $s_i=\frac{E_{i+1}-E_i}{\langle E_{i+1}-E_i\rangle_\mathrm{ER}}$.

Yet, the distribution of adjacent level spacings alone does not reveal the transition towards quantum chaos. To overcome this issue, the level spacing ratio was introduced~\cite{Huse_SpacingRatio_MBL, PalHuseHeisenbergMBL, sherringtonKirkpatrickMBL}
\begin{equation}
    r_i=\frac{\min{(s_{i+1},s_i)}}{\max{(s_{i+1},s_i)}}\,,
\end{equation}
which is robust against variations in the local density of states. 
It captures the emerging chaotic behavior by its average value $\langle r\rangle$, approaching $r_\mathrm{WD}\approx 0.523$ in the thermodynamic limit. Convergence towards either the Poissonian $r_\mathrm{Poi}\approx0.386$, or a lower value signals the breakdown of quantum chaos. The level spacing ratio distributions read 
\begin{equation}    
       \tilde P_\mathrm{Poi}(r)\propto\frac{1}{(r+1)^2},\quad \tilde P_\mathrm{WD}(r)\propto\frac{r+r^2}{(1+r+r^2)^{5/2}}\,.
\end{equation}
 Additional spectral diagnostics via these statistics are provided in App.~\ref{app:P_rStat}.
 
We perform numerical simulations to investigate the onset of quantum chaos from the perturbatively small $g = 0.01J$ to the fully polarized limit $g = 5J$. The connectance is varied from low to high, $\tilde M\in[0.07,\,0.98]$.
 To study the spacing distribution in a spectrally resolved way, we analyze ten uniformly spaced energy windows in the \emph{unfolded} spectrum. The unfolding procedure~\cite{ProzenVidamerSFF_MBLChaos, Bertrand_Unfolding_strategies} facilitates the characterization of the local structure of level spacing distribution. It relies on scaling out the local density of states
$\rho(E)=\left\langle\sum_i \delta(E-E_i)\right\rangle_E$, computed numerically for each graph realization. Here, the average $\langle\dots\rangle_E$ runs over the given energy windows. The cumulative distribution function is obtained by integrating the density, $f(E)=\int_{-\infty}^{E}\mathrm dE^\prime\,\rho(E^\prime)$. The unfolded energies $\tilde E_i=f(E_i)$ are then mapped by fitting a fifth-order polynomial to the cumulative distribution function.

The spectrally resolved analysis reveals the effects of the discrete-valued disorder. In particular, the total bandwidth of $H_P$ scales linearly with $J$ having $\frac{2J}{M}$ spacings. The number of degenerate eigenstates at $g=0$ with spins pointing opposite ways in closed loops grows exponentially with $L$. With finite $g$, the spectrum inherits this behavior to some extent.
  These frustrated spectral regions reside with high precision in the upper half of the spectrum, $\tilde E\gtrsim J/2$. We observe that the related features appear mainly at small spacing values, but do not change the overall structure of the statistics (see App.~\ref{App: energy_spectrum}).
 For these reasons, the spectral analysis is carried out in the lower half of the unfolded spectrum in the energy window of $[0.3J,\, 0.4J]$. This regime is sufficiently below the frustrated upper half but close enough to the middle to test the bulk spectral properties. 
 
As shown in Fig.~\ref{fig: LStat_examples} $(e)$, for intermediate $0.5\,J\lesssim g\lesssim 4\,J$ and for $1/L\ll\tilde M<1$ the level spacing distributions converge towards the WD distribution with increasing $L$. On the other hand, for small values of $g=0.1\,J$, level repulsion is weaker, but it still increases with the system size as demonstrated in Fig.~\ref{fig: LStat_examples} $(h)$. For small sizes, the distributions are more concentrated around zero than the Poissonian one, but for $L=15$ they already reach in between the Poissonian and WD shapes. The results indicate no slowing down of this tendency, thus it is expected that the distributions eventually reach the WD shape for larger but numerically not achievable values of $L$.
The effect of small transverse fields on the quantum chaotic signatures has been studied in various spin models including, for instance, Ising ladders, tilted and two-dimensional Ising models~\cite{Rigol_Weakg_ETH_2016,KareljanWeak_g_2020,Bastinello_2022,Hamazaki_Weakg_2022,Harkema_2024}. 

In the present case, this behavior can be captured using first-order perturbation theory at $g=0$. The overlap is computed between two non-degenerate eigenstates $\lvert k_1\rangle,\,\lvert k_2\rangle$, separated by unit Hamming distance around the middle of the energy spectrum,
\begin{equation}\label{eq:eq_prturb_1_z}
    T_{k_1,k_2}\approx\frac{g}{L}\frac{1}{\Delta E_{12}},
\end{equation}
where $\Delta E_{12}\sim\frac{\sqrt L}{M}$ is the typical scale of the energy difference (see App.~\ref{App: Onsite_Energies}). Requiring that $T_{k_1,k_2}\sim \mathcal{O}(1)$ in the thermodynamic limit sets the threshold of the transverse field $g\gtrsim L^{-1/2}J$. Below this scale, the energy contributions are smaller than the mean level spacing between non-degenerate eigenstates differing by one spin-flip. The transverse field provides only small corrections, implying trivially localized behavior. Thus, the spacing distribution gets concentrated around zero more than the Poissonian distribution with a Shnirelman peak~\cite{Schnierelman_peaks_KickedRotor, Schnierelman_peaks_QMBilliard}, as displayed in Fig.~\ref{fig: LStat_examples}  $(h)$. Similar behavior was reported in Heisenberg models with strongly correlated random longitudinal fields~\cite{Schnierelman_peaks_SpinSystems, CorrealtedDisorder_Shnirelman_peak}. 

\begin{figure*}[t!]
   \includegraphics[width=1\linewidth,trim={5cm 2cm 45cm 0},clip]{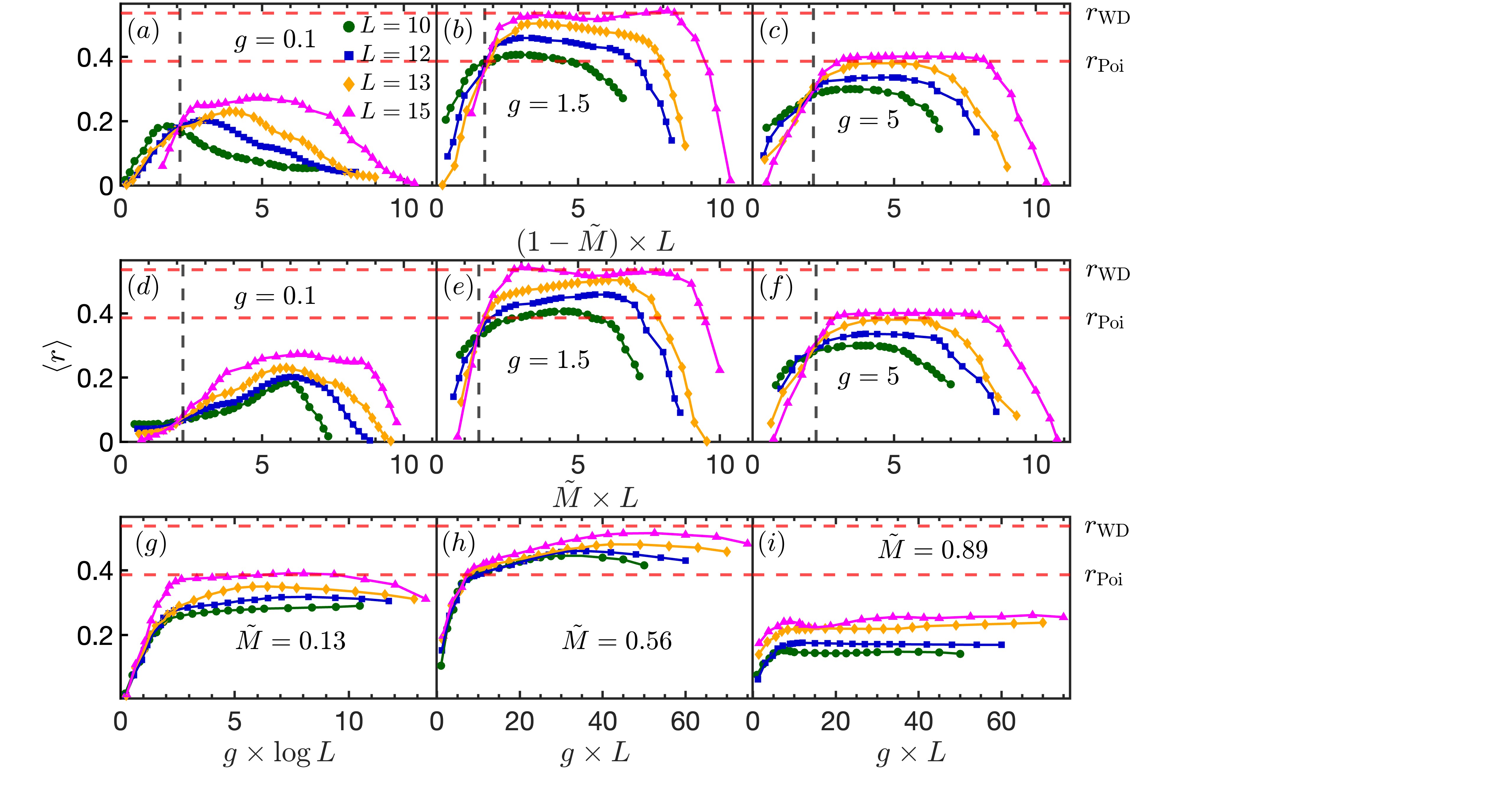}   
    \caption{
    Level spacing ratio as a function of $\tilde M$ and $g$ for $L=10,12,13,15$ in the energy window $[0.3\,J,\,0.4\,J]$ of the unfolded spectrum.
    Panels $(a)-(c)$: Level spacing ratio as a function of $(1-\tilde M)\times L$ for $g=0.1J,\,1.5J,\,5J$, exhibiting the transition to the fully connected limit around $(1-\tilde M)\times L\approx2$.
    Panels $(d)-(f)$: Same $g$ values predicting the breakdown of quantum chaos around the disconnected limit at $\tilde M\times L\approx1.8$. The vertical dashed lines denote the obtained transition points around the disconnected and LMG limits.
    Panels $(g)-(i)$: Level spacing ratio as a function of the rescaled $g$ for low, intermediate, and high $\tilde M$ values. Red dashed lines denote the Poissonian and WD values of $\langle r\rangle$ in all the panels.} \label{fig:ratioavg_examples}
\end{figure*}

However, with increasing $L$, the spectrum escapes this perturbative regime at fixed $g$ values and develops chaotic behavior, as displayed by Fig.~\ref{fig: LStat_examples} and Fig.~\ref{fig:ratioavg_examples} $(h)$.
To understand the non-trivial breakdown of quantum chaos, we examine the condition for the instantaneous eigenstates of $H[g]$ to have finite overlaps with computational basis states that are separated by $\sim O(\sqrt L)$ Hamming distances. Satisfying this condition implies the long-range delocalization of the eigenstates and thus the onset of quantum chaotic behavior.
As shown in App.~\ref{App: Perturbative_Calc},  perturbative corrections of order $n\sim\sqrt L$ order in the forward scattering approximation~\cite{Schardicchio_forward_scattering, Schardicchio_Bethe} result in an amplitude,
\begin{equation}\label{eq:Large_Order_z}
\begin{split}
    &T_{k_1,k_2}\propto \left(\frac{g}{L}\right)^n\sum_{m_1,m_2,\cdots,m_{n-1}} (H_x)_{k_1,m_1}\cdots(H_x)_{k_1,k_2}\\
    &\times\left(\Delta E_{k_1,k_2}\prod_{i=1}^{n-1}\Delta E_{k_1,m_i}\right)^{-1}\sim \left(\frac{gL}{J}\right)^n,
    \end{split}
\end{equation}
posing the lower boundary $g\gtrsim L^{-1}J$. For these $g$ values, exponential suppression of eigenstate delocalization breaks down and quantum chaos sets in as a universal function of $g\times L$.
 This is verified by the scaling collapse of the $\langle r\rangle$ values in Fig.~\ref{fig:ratioavg_examples} $(h)$.
 Furthermore, the dominance of long-range spreading of the eigenstates implies that no mobility edge exists, in agreement with Fig.~\ref{fig:ratioavg_examples} $(g)-(i)$.
 By contrast, in the single-particle case~\cite{Anderson_original_1, Anderson_original_2, Anderson_original_3},
the critical values of the hopping amplitudes are of the same order as the mean level spacing. The present analysis predicts behavior similar to that of approximate methods in Refs.~\cite{Rigol_IntBreak_g_c, Levitov_IntBreak}. The analysis of the $\langle r\rangle$ values for $g=0.1\,J$ and $g=1.5\,J$, exhibits the same behavior.
 As shown in Fig.~\ref{fig:ratioavg_examples} $(a),\,(b),\,(d)$ and $(e)$, convergence for intermediate $\tilde M$ is observed towards the chaotic value $r_\mathrm{WD}\approx0.52$. 
 In agreement with the results of the spacing distribution, convergence is the fastest for intermediate $g$ values.
 
    \begin{figure*}[t!]
    \includegraphics[scale=0.1,trim={55 605 2.4cm 203},clip]{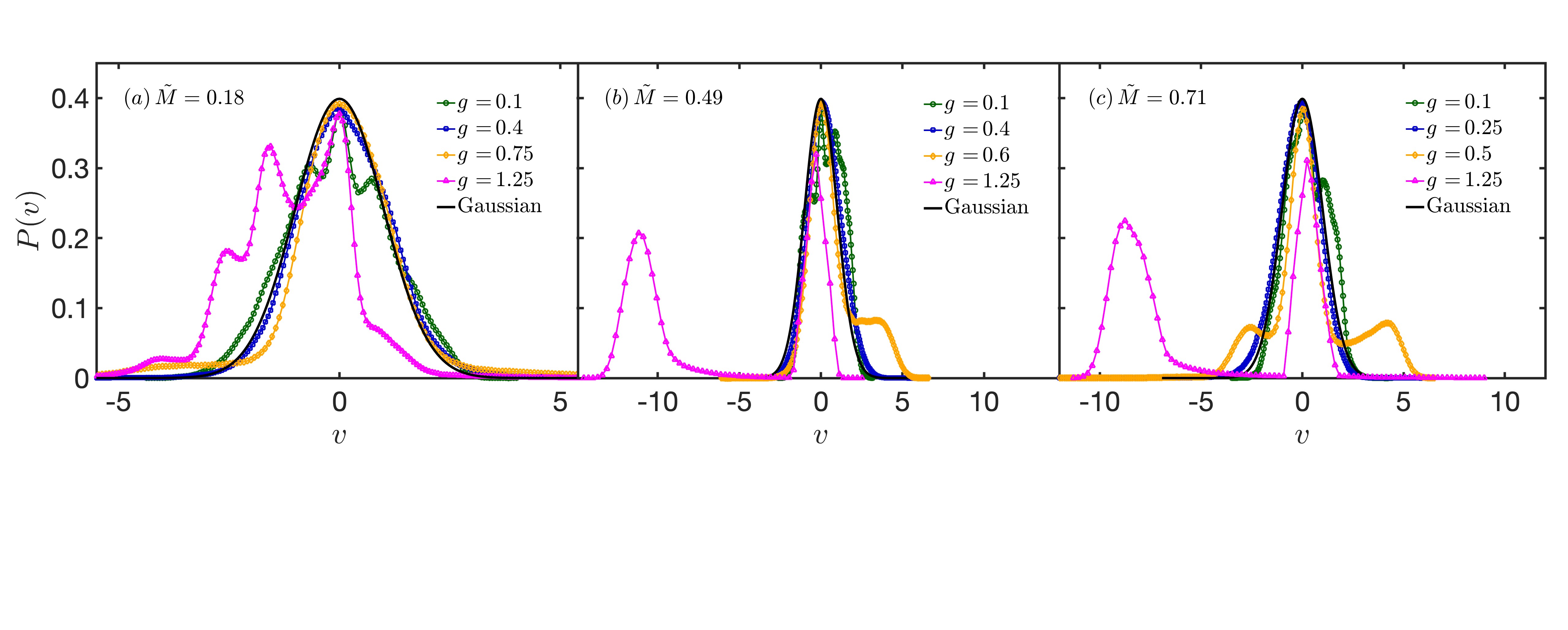} 
    \caption{Level velocity distributions for $L = 14$ scaled to unit variance, compared with a standard Gaussian distribution. The figure shows results for $(a)$ $\tilde M = 0.18$, $(b)$ $\tilde M = 0.49$, and $(c)$ $\tilde M = 0.71$. Deviations from the Gaussian curve is observed as the transverse field $g$ increases. This happens for smaller $g$ values as $\tilde M$ increases.} 
    \label{fig4}
\end{figure*}

\subsection{Sparsely and densely connected graphs}

 Next, we turn to the cases of low connectance with a growing number of disconnected subnetworks, $\tilde M\lesssim 2/L$, and the LMG limit with $1-\tilde M\gtrsim 2/L$. 
 
 
 The ER graph falls apart to disconnected subnetworks of size $\sim\log L$~\cite{Erdos_Renyi_Graph_1, Erdos_Renyi_Graph_2} when the connectivity reaches the order of the number of vertices.
 It simplifies to $\tilde M=\frac{1}{L}$ in the thermodynamic limit. 
 The agreement with the WD statistics naturally breaks down due to these graph-theoretical features. The energy spectrum splits up to the sum of the $\sim L/\log L$ completely uncorrelated spectra with $\sim2^{\log L}$ energy levels.
  The number of the subnetworks is exponentially larger than their sizes, and therefore the corresponding independent energy spectra can be characterized by their average spectral properties in the $L\rightarrow\infty$ limit.
  Following the lines of Ref.~\cite{Tindall2022_QMconnectedNeworks} we approximate these disconnected spin systems with large $\log L$ collective spins. As the total spectrum is given by all the combinations of the sum of individual levels of the disconnected parts, the probability of having exact degeneracies becomes finite.
   This is in contrast with the Poissonian case of independent identical levels.
  Thus, the corresponding spacing distribution acquires a larger weight than the Poissonian one around zero spacings.
  However, the disconnected regime survives only when the rescaled connectance is kept constant, $\tilde M\, L\lesssim2$.
  For fixed $\tilde M$, the system inherently escapes this regime and restores quantum chaos.

  These features are demonstrated in Fig.~\ref{fig: LStat_examples} $(a),\,(d)$ and $(g)$. In particular, the corresponding distributions only exhibit an early stage of convergence towards the WD shape for the reachable system sizes. For intermediate $g$ the spacing distributions already acquire an intermediate shape between the Poissonian and WD ones, as shown in Fig.~\ref{fig: LStat_examples} $(d)$. For small and large $g$ values chaotic signatures are weaker for the achievable system sizes, as demonstrated in Fig.~\ref{fig: LStat_examples} $(a),(g)$. Yet, level repulsion increases with the system size, as the corresponding distributions get less and less concentrated around small spacing values. For larger values of $L$, these curves are expected to reach the WD distribution.
  
  The perturbative argument of the onset of chaos in Eq.~\eqref{eq:Large_Order_z} naturally carries on to the disconnected subnetworks. This implies that the universal scale of $g$ is governed by the small subgraph sizes separately. As demonstrated in Fig.~\ref{fig:ratioavg_examples} $(g)$, this leads to the scaling collapse as a function of $g\times \log L$.

 Remarkably, the level spacing ratio captures the connected-disconnected transition point.
 As shown in Fig.~\ref{fig:ratioavg_examples} $(d)-(f)$, at fixed $g$ values, the corresponding ratio curves cross each other at approximately the same value of the rescaled connectance, $\tilde ML\approx1.8$. This slight deviation from the theoretical value is the consequence of finite-size effects, which are especially relevant in the disconnected subgraphs. This further demonstrates that the distributions for arbitrary small $\tilde M$ will converge towards the WD distribution for large enough system sizes.
 Additionally, the transition point shifts to higher values for limiting $g$ due to additional localization effects. These findings are demonstrated in Fig.~\ref{fig:ratioavg_examples} $(d)-(f)$ for small, intermediate, and large $g$ values.

 Turning to the opposite limit, the scale $1-\tilde M\gtrsim 2/L$ naturally leads to the vanishing of the relative variances and correlations according to Eq.~\eqref{eq:relative_correlators},
 \begin{align}    
    &\frac{\sqrt{\langle \delta E^2_k\rangle_\mathrm{ER}}}{\langle E_k\rangle_\mathrm{ER}}\sim L^{-1/2}\,,\\
    &\frac{\sqrt{\langle\delta E_k\delta E_{k+r}\rangle_\mathrm{ER}}}{\langle E_k\rangle_\mathrm{ER}}\sim L^{-1/2}\sqrt{1+6\frac{r}{L}\left(\frac{r}{L}-1\right)}\,.
 \end{align}
 This implies that with increasing $L$, on-site energy distributions get concentrated around the average $\langle E_k\rangle_\mathrm{ER}=\tilde M\frac{S^2_k-L}{M}$. Thus, the spectrum becomes identical to that of the LMG model up to a vanishingly small correction due to the $\tilde M$ prefactor and with a relative error of
 \begin{equation}\label{eq: LMG_Error}
     \frac{E_k-E^\mathrm{LMG}_k}{E^\mathrm{LMG}_k}\sim\mathcal O(L^{{-1/2}})\,.
 \end{equation}
 Note that Eq.~\eqref{eq: LMG_Error} is in agreement with a more general exact relation reported in Ref.~\cite{Tindall_Graph_Theory_FreeEnergy}.
 Here, $E^\mathrm{LMG}_k=\frac{S^2_k-L}{M}$ denotes the energy of the LMG model at $g=0$.
 This error originates from the on-site variances rather than from the $\tilde M$ prefactor.  Involving $H_x$ to the Hamiltonian will result in the same relative error compared to the transverse field LMG model~\cite{FCQTFIM_MF}. 
 
 Close to the fully connected limit, quantum chaos breaks around the complementary scale of the disconnected limit, $1-\tilde M\approx 2/L$.  The transition also becomes sharp, as captured by the level spacing ratios. In Fig.~\ref{fig:ratioavg_examples} $(a),\,(b)$ a transition point as a function of $(1-\tilde M)\times L$ is displayed. The crossover sets in at slightly larger values than $1-\tilde M\gtrsim2/L$ for limiting $g$ values. 
 
 Similar to the regime of the intermediate values of $\tilde M$, the level spacing distributions exhibit an initial stage of the convergence towards the WD shape as they develop stronger and stronger level repulsion as $L$ grows, as shown in Fig.~\ref{fig: LStat_examples} $(f)$, and $(i)$. For intermediate $g$ in panel $(f)$ the distributions reach in between the Poissonian and WD shapes, while for small $g$ in panel $(i)$ only the tendency can be observed, that the distributions are getting less concentrated around zero with increasing $L$.
 These predictions are further validated by the sharp transition points in the level spacing ratios as a function of the rescaled connectance, $(1-\tilde M)\times L$ presented in Fig.~\ref{fig:ratioavg_examples} $(a)-(c)$.
 
 These results show the main difference between fixing $\tilde M$ or $(1-\tilde M)\times L$. The former leads to a convergence towards the regime of intermediate $\tilde M$ and quantum chaos. In the latter case, the system stays near the LMG limit with vanishing on-site fluctuations upon increasing $L$. This explains well the decreasing behavior of the $\langle r\rangle$ values, as demonstrated Fig.~\ref{fig:ratioavg_examples} $(a)-(c)$.

\begin{figure}[t!]
    \includegraphics[scale=0.12,trim={225 90 103cm 0},clip]{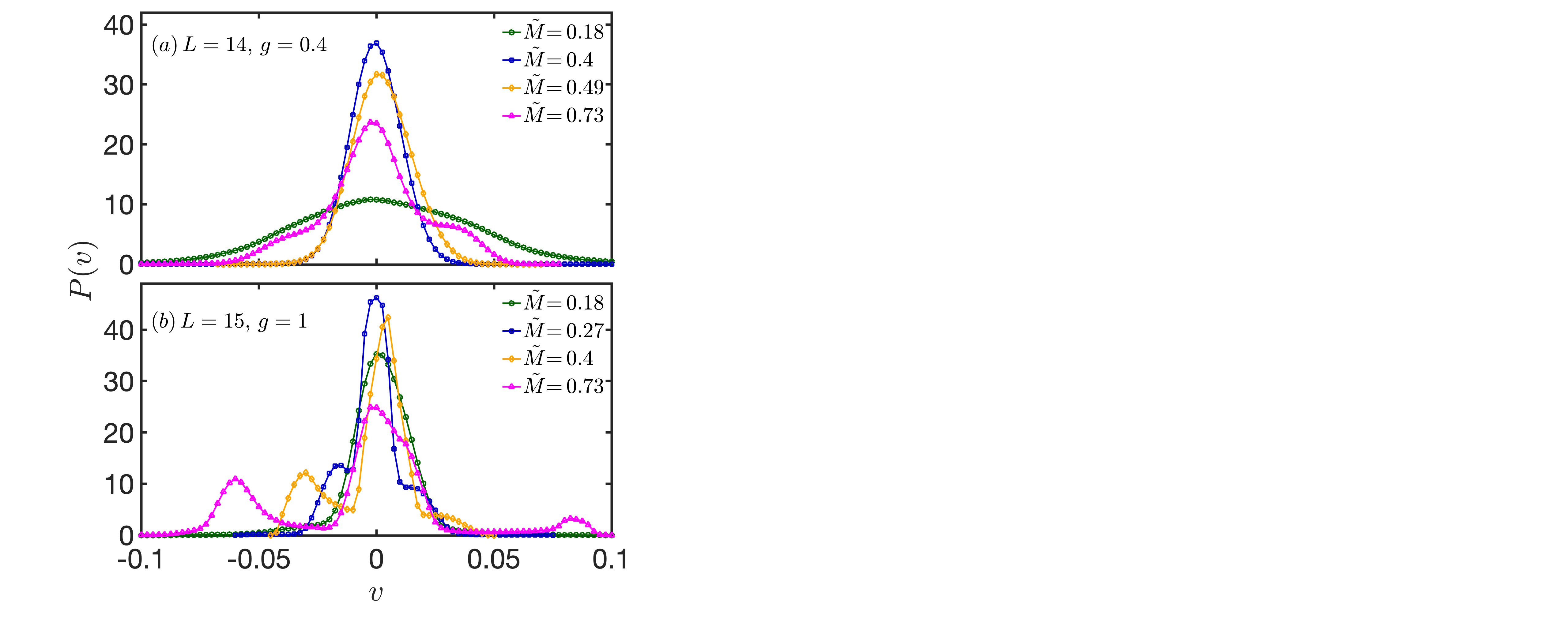} 
    \caption{Level velocity distributions for low to high connectances $\tilde M = [0.18, 0.73]$. The figure shows results for $(a)$ $L=14, g = 0.4\,J$, and $(b)$ $L=15, g = J$. For the  larger $g$ value, the velocity statistics break down earlier with increasing $\tilde M$.} 
    \label{fig5}
\end{figure}
\subsection{Strongly polarized limit}
 Finally, the strongly polarized limit $ g\gg J$ is discussed, which provides similar thresholds for the breakdown of the quantum chaotic behavior. Large $g$ values lead to localization in the $X$ polarized basis. Due to the degeneracies in the $X$ basis, the spacing distribution clusters around zero more pronouncedly than the Poissonian distribution.
 Near the LMG limit, the rescaled transition connectance $(1-\tilde M)\times L$ increases with $g$ entering the intermediate regime. For fixed $\tilde M$, the increasing $L$ restores quantum chaos, as observed in Fig.~\ref{fig: LStat_examples} $(c)$ and Fig.~\ref{fig:ratioavg_examples} $(i)$. In particular, these figures demonstrate, how level repulsion increases with the system size by the spacing distributions getting less peaked around zero and by the growth of the spacing ratios. Even though, for the reachable system sizes level repulsion is weak, proper convergence towards the WD distribution is expected for larger values of $L$ based on the transition point in terms of the spacing ratios, Fig.~\ref{fig:ratioavg_examples} $(c)$.
 In the connected $2/L\ll\tilde M< 1-2/L$, and disconnected $\tilde M\lesssim2/L$ regimes,  distinct scalings are found for the onset of quantum chaos. For intermediate $\tilde M$, first order perturbative approximation of the overlap of two $X$ polarized eigenstates, $\lvert x_1\rangle,\,\lvert x_2\rangle$ scales as
 \begin{equation}
     T_{x_1,x_2} \approx \frac{LJ}{Mg} \sim \frac{J}{Lg} \,.
 \end{equation}
 Here, the two eigenstates are separated by two spin-flips around the middle of the spectrum,
 which fixes the threshold as $g\lesssim L^{-1}J$.  Above this threshold the short-range delocalization of the eigenstates of $H[g]$ is suppressed. It means that the overlaps between the instantaneous eigenstates and the $X$ basis states that are separated by finite, $\mathcal O(1)$ Hamming distance are negligibly small.

 By contrast, the onset of quantum chaos near the strongly polarized limit is better encapsulated by long-range delocalization.
 Employing again the forward scattering approximation of order $n\sim\sqrt L$, we obtain
 \begin{equation}\label{eq:large_order_H_x}
 \begin{split}
     &T_{x_1,x_2}\propto \sum_{m_1,\cdots,m_{n-1}} (H_P)_{x_1,m_1}\cdots(H_P)_{m_{n-1},x_2}\\
    &\times M^{-n}\left(\Delta E_{x_1,x_2}\prod_{i=1}^{n-1}\Delta E_{x_1,m_i}\right)^{-1}\sim\left(\frac{\sqrt LJ}{g}\right)^n,
    \end{split}
 \end{equation}
 leading to the threshold of $g\lesssim L^{1/2}J$ (see App.~\ref{App: Perturbative_Calc}). Therefore, strong transverse field terms cannot challenge quantum chaos beyond suppressing short-range hybridization. Increasing $L$ facilitates the restoration of quantum chaos.
 The whole procedure can be repeated in the disconnected regime by replacing $L$ with $\log L$. The uncorrelated subnetworks are separately governed by the strong transverse fields. This implies that the threshold scales with the subnetwork sizes, $g\sim\sqrt{\log L} J$. 
 The convergence of the level spacing distribution and the spacing ratios are presented in Fig.~\ref{fig: LStat_examples} $(a),(b)$ and Fig.~\ref{fig:ratioavg_examples} $(g),\,(h)$. This convergence slows down near the disconnected regime. This implies a similar weak feature, namely, that the convergence towards the WD distribution and spacing ratio value is only observed in an early stage due to the numerical restrictions on reaching larger system sizes.

\section{Level velocity analysis}\label{sec:VelStat}

\noindent This section addresses the chaotic signatures of the spectrum from the point of view of energy level dynamics~\cite{Yukawa_1985, Pechuka_1983, Haake_QM_Signatures_2001}.
The parametric variations of energy levels of a Hamiltonian $H[g]$ capture non-equilibrium phenomena induced by external driving potentials in disordered systems. The corresponding derivative of the energy levels is referred to as ``level velocity'' $v_{n_g}=\frac{\mathrm dE_n}{\mathrm dg}$. This quantity effectively measures the sensitivity of the spectrum against external perturbations.
Prominent examples are the fluctuations of the microscopical conductance in the presence of time-dependent magnetic fields~\cite{Magnetic_field_1, Magnetic_field_2}. 
Further research has focused on level dynamics in disorder Heisenberg models and one-dimensional quantum systems~\cite{THOULESS_original, Edwards_Vel_1972, Akkermans_1992_sensitivity}. 
Universal velocity correlations were revealed in disordered hopping models with changing boundaries. Analytical velocity statistics were derived in the localized phase of driven Anderson models~\cite{Fyodorov_VelStat1, Fyodorov_VelStat2, Simmons_VelCorr1, Simmons_VelCorr2}.

\begin{figure*}[t!]
    \includegraphics[scale=0.10,trim={25 605 2.4cm 50},clip]{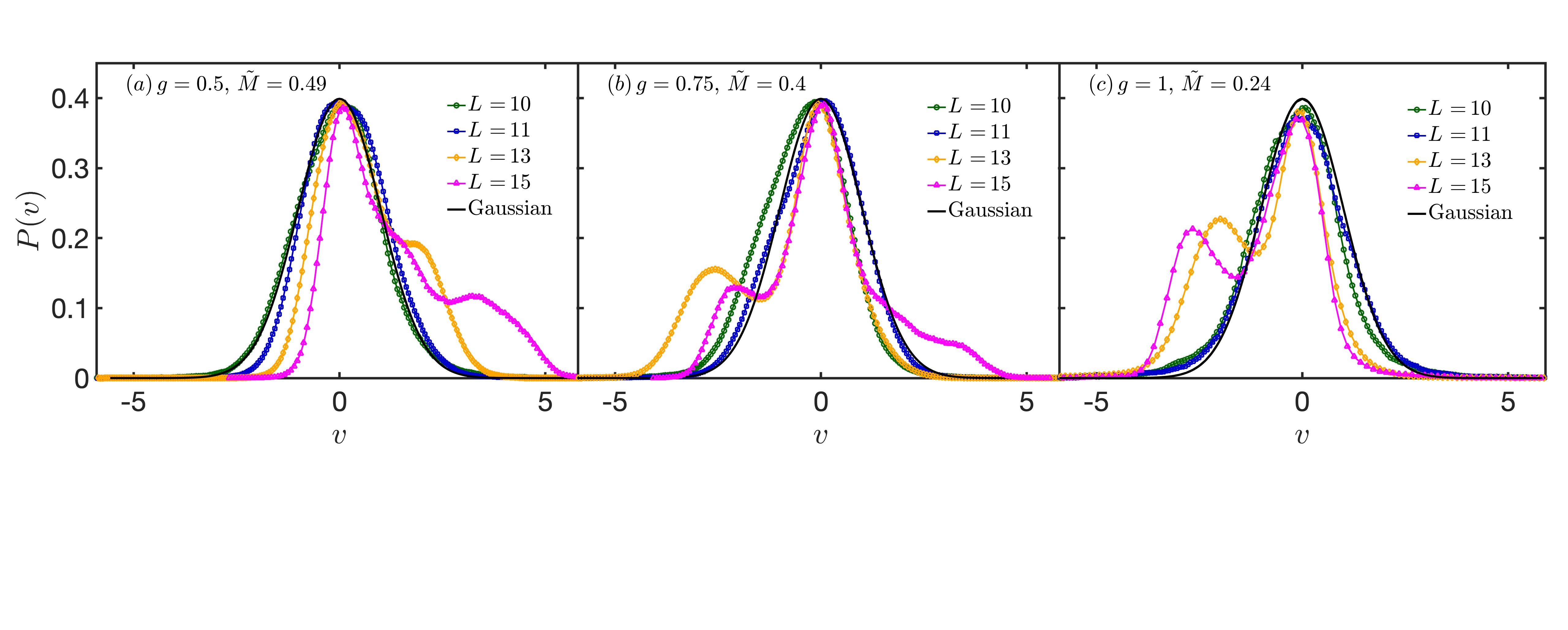} 
    \caption{Level velocity statistics scaled to unit variance compared to standard Gaussian distribution. The figure shows results for $(a)$  $g=0.5\,J,\tilde M = 0.49$, $(b)$ $g = 0.75\,J,\tilde M = 0.4$, $(c)$  $g = J,\tilde M = 0.24$, for $L=10,11,13,15$. Increasing $L$ leads to strong deviations compared to the the Gaussian shape. 
    This is observed for $L =13, 15$ whereas $L=10,11$ approximately follows the Gaussian curve. Furthermore, deviations appear earlier for increasing $g$ and $\tilde M$.} 
    \label{fig6}
\end{figure*}

Similar to the spacing distribution, level velocity 
\begin{equation}
    v_{n_g}=\frac{\mathrm dE_{n_g}}{\mathrm dg} = \frac{1}{L} \bra{n_g} {H}_x \ket{n_g}, \label{eq20}
\end{equation}
is used as a local characterization of the spectral properties. Here, $\ket{n_g}$ refers to the eigenstate of $H[g]$ in Eq.~\eqref{eq2}. To this end, we probe chaotic signatures via $v_{n_g}$ as a function of $g$ and $\tilde M$. 

The level velocity provides valuable insights into the statistical properties of both the eigenvalues and eigenstates. To capture the local spectral properties, the unfolding method is applied similarly to level spacing,
\begin{equation}
    v_{n_g}\rightarrow \tilde v_{n_g} = \frac{d \tilde E_{n_g}}{dg} = \rho(E_{n_g})v_{n_g}, \label{eq21}
\end{equation}
where $\tilde v_{n_g}$ represents the unfolded level velocities. The eigenstates of $H[g]$ can be expanded as
\begin{equation}
    v_{n_g} = \frac{1}{L} \bra{n_g} {H}_x \ket{n_g} = \frac{1}{L} \sum_{k,l} C_l^* C_k \bra{l} {H}_x \ket{k}, \label{eq22}
\end{equation}
where $\ket{k}, \ket{l}$ are the eigenstates of $H_P$, with amplitudes $C_k$, $C_l$. The non-zero terms in Eq.~\eqref{eq22} occur when $\ket{k}$ and $\ket{l}$ differ by only one spin, leading to
\begin{equation}
    v_{n_g} = \frac{1}{L} \sum_{k,l} C_l^* C_k \delta(\Delta S_{kl}- 1) \label{eq23},
\end{equation}
where $\Delta S_{kl} = \sum_i |\sigma^{(k)}_i-\sigma^{(l)}_i|$ and $\sigma^{(k)}_i=\pm1/2$ denotes the $i^{\text{th}}$ spin $Z$ value of the $k^{\text{th}}$ spin configuration. Based on the level spacing analysis, eigenstates are captured by RMT up to high precision. This leads to approximately independent and extended amplitudes $C_k,\,C_l\sim\frac{1}{\sqrt{2^L}}$. 
Thus, Eq.~\eqref{eq23} leads to Gaussian velocity statistics in quantum chaotic spectral regions, which is in agreement with previous findings ~\cite{PhysRevB.99.224202, PhysRevB.107.014206}.
However, its deviation from the Gaussian shape does not necessarily capture the integrable regimes. In contrast to the spacing distribution, $v_{n_g}$ is governed by short-ranged eigenstate delocalization for small $g$. In Eq.~\eqref{eq23}, $C_k,\,C_l$ spread uniformly over the threshold $g\gtrsim L^{-1/2}$.

The breakdown of quantum chaos in terms of the deviations of the level velocity statistics from the Gaussian distribution can be understood via first-order perturbation theory. For small $g$ values, $H_x$ can be treated as a perturbation to  $H_P$.  In this limit (see App. \ref{App: velocity_analytics})
\begin{equation}
    v_{n_g} \sim g \sqrt{\tilde M}. \label{eq24_1}
\end{equation}
Note that the estimation of these scales is not affected by the density of states, therefore these predictions remain valid also for the unfolded velocity, $\tilde v_{n_g}$.
In the strongly polarized limit, Eq.~\eqref{eq20} is governed by the eigenstates of $H_x$ in leading order around the middle of the spectrum yielding (see App.~\ref{App: velocity_analytics}),
\begin{equation}
    v_{n_g} \sim \frac{1}{\sqrt{L}}\,. \label{eq25_1}
\end{equation}
Based on numerical results the intermediate regime between the strongly polarized and small $g$ limits is narrow enough, allowing us to identify the breakdown by matching the two scales. 
Comparing Eq.~\eqref{eq24_1} and Eq.~\eqref{eq25_1} leads to the threshold scale
\begin{equation}
    g \propto \frac{1}{\sqrt{\tilde M L}}, \label{eq26_1}
\end{equation}
where $v_{n_g}$ breaks down. It further validates the first-order perturbation approach since, for large $L$, low-order perturbation corrections converge with the breakdown scale.

The velocity statistics breaks down near the perturbatively small and strongly polarized $g$ limits at fixed $\tilde M$ and $L$, as shown in Fig.~\ref{fig4}.
The dependence on the connectance exhibits different behavior in the disconnected limit $\tilde M\sim1/L$, as the Gaussian shape persists with reasonably increased variances.
While the spacing distribution shows strong deviations from the WD statistics in the disconnected regime, the velocity statistics is a proper candidate to study chaotic signatures of the individual subnetworks.
In particular, Gaussian velocity statistics survive separately in these disconnected parts, which sum up as independent random variables. This results in a final normal distribution in line with the central limit theorem.
However, the densely connected regime exhibits the same behavior as in the preceding section, where Gaussian shape and quantum chaos break down. These results are presented in Fig.~\ref{fig5}. 

Furthermore, in the disconnected regime, the upper threshold of $g$ also increases in contrast to the spacing analysis. This property is well explained by the scaling in Eq.~\eqref{eq26_1} as the crossover connectance $\tilde M\sim1/L$ implies a size-independent threshold of $g$. Close to the fully connected limit, quantum chaos survives only in a narrower window of $g$, as similarly revealed by the spacing analysis.

In contrast to the level spacing distribution, increasing $L$ leads to the breakdown of the chaotic character.
Near the strongly polarized limit, considering $H_P$ as a small perturbation, the velocity reads
\begin{equation}
    v_{n_g}=\frac{1}{L}\sum_{x_l,x_k}\,C^*_{x_l}C_{x_k}\left\langle x_l\left\lvert H_x\right\rvert x_k\right \rangle=\frac{1}{L}\sum_{x_k}\,\lvert C_{x_k}\rvert^2\,,
\end{equation}
with $\lvert x_l\rangle,\lvert x_k\rangle$ denoting the eigenstates of $H_x$.
Here, the velocity statistics are governed by the amplitudes of the $X$ polarized eigenstates. The spreading of these eigenstates becomes strong enough when $g$ is sufficiently below the long-range delocalization threshold. The $C_{x_k}$ amplitudes delocalize in the $X$ basis satisfying the conditions of Gaussian velocity statistics at the scale $g\sim L^{-1/2}$.
The deviations from the Gaussian shape against increasing the system size is illustrated in Fig.~\ref{fig6}. For smaller sizes $ L = 10, 11$, the distribution still follows precisely the Gaussian shape, whereas the statistics for higher ones with $L = 13, 15$ show clear deviations.

\begin{figure}[t!]
    \includegraphics[width=1\linewidth,trim={5cm 5 45cm 0},clip]{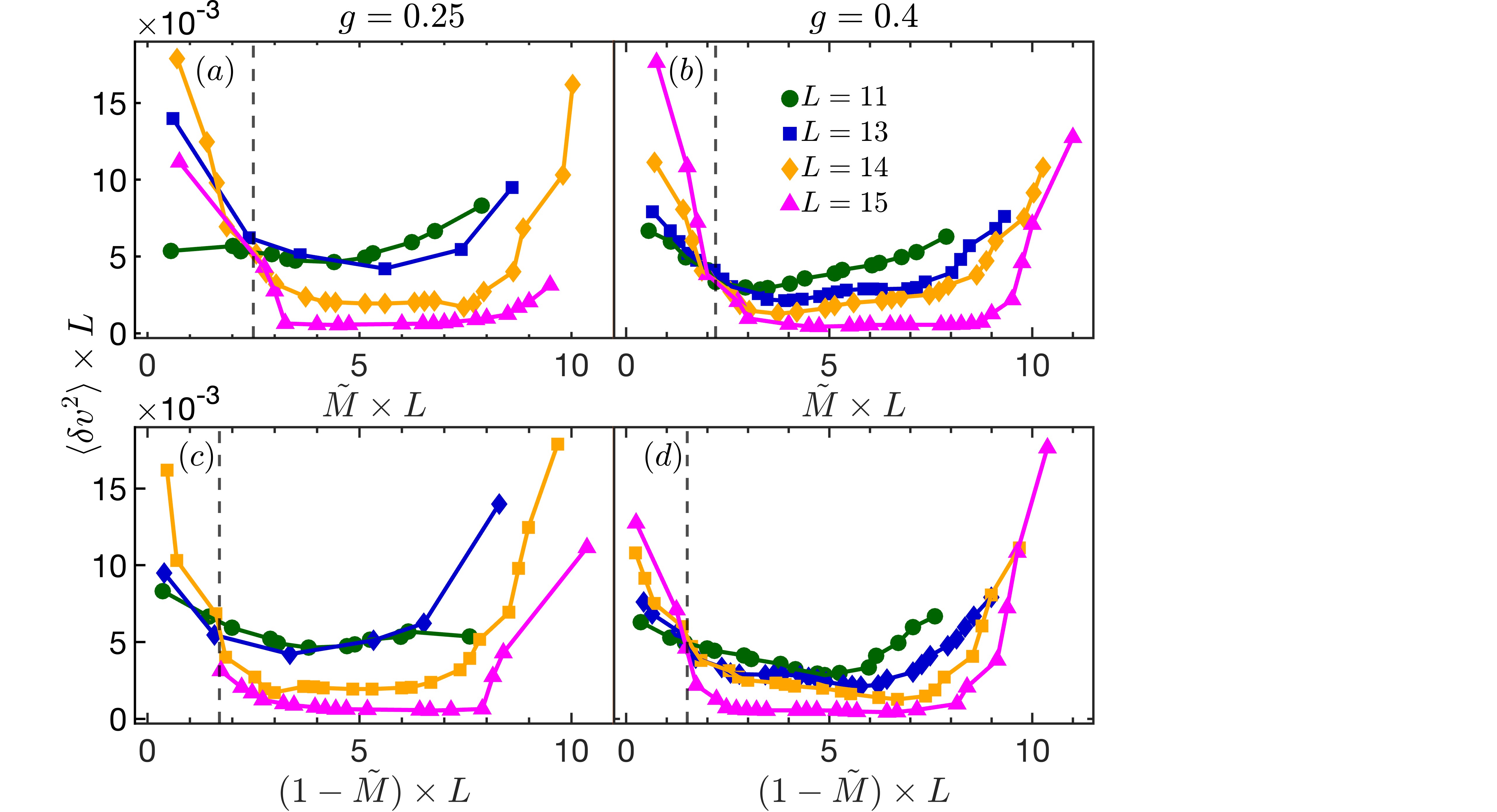} 
    \caption{Rescaled level velocity variances for $g\lesssim 1.25\,J$ values where the Gaussian shape persists.
    Panels $(a)$ and $(b)$: Transition point around the disconnected regime as a function at $\tilde M\times L\approx1.8$ for $g=0.25\,J,\,0.4\,J$.
    Panels $(c)$ and $(d)$: Transition towards the LMG limit at $(1-\tilde M)\times L\approx1.6$ for $g=0.25\,J,\,0.4\,J$. The transition points to the disconnected and fully connected limits are highlighted by the vertical dashed lines.
    }\label{fig:VelVar}
\end{figure}

The variance of the level velocities is investigated both in the connected and the disconnected regimes around the middle of the spectrum. Using Eq.~\eqref{eq23} the approximate RMT description of the eigenstates, the variance reads as
\begin{equation}
    \left\langle \delta v^2 \right\rangle \sim \frac{1}{L} \sum_{k,l} \left\langle (C_l^* C_k)^2 \right\rangle \delta(\Delta S_{kl} - 1) \label{eq24}.
\end{equation}
Approximating $C_k,\,C_l\sim1/\sqrt{2^L}$ Eq.~\eqref{eq24} becomes
\begin{equation}
    \left\langle \delta v^2 \right\rangle \sim \frac{1}{L}  \sum_{k,l}  \frac{1}{2^{2L}} \delta(\Delta S_{kl} - 1) \label{eq25}.
\end{equation}
However, there are a total of $2^L L$ possible ways to obtain non-zero delta functions. Hence, $\left\langle \delta v^2 \right\rangle  \sim \frac{2^L L}{L}\frac{1}{2^{2L}}\sim \frac{1}{2^L}$, in the chaotic region. This analysis predicts a fast decrease in $\left\langle \delta v^2 \right\rangle$ with respect to the system size $L$.
The corresponding level velocity statistics are depicted in App.~\ref{App: velocity_analytics}.

In the sparsely connected regime, $\langle\delta v^2\rangle$ exhibits a different scaling. Although the chaotic behavior of the velocities survives in the disconnected regime, $\langle\delta v^2\rangle$ still captures the corresponding transition point.
The variance exhibits a sudden growth around $\tilde M\approx2/L$ due to the independence of the chaotic subnetworks. In particular, the level velocities of the subnetworks add up as independent Gaussian random variables, eventually leading to a larger system size scale of $\langle\delta v^2\rangle$,
\begin{equation}
\begin{split}
    \left\langle \delta v^2 \right\rangle & \sim \frac{1}{L} \sum_{\mathrm{subgraphs}} \sum_{k,l} \left({\frac{1}{{2^{\log L}}}}\right)^2 \delta(\Delta S_{kl} - 1) \\ & \sim  \frac{1}{L} \sum_{\mathrm{subgraphs}} \frac{1}{2^{2\log L}} 2^{\log L} \log L \\ & \sim  \frac{1}{L}  \frac{\log L}{2^{\log L}} \frac{L}{\log L} = L^{-\log 2}\label{eq26}.
\end{split}
\end{equation}
The first summation is performed over the $L/\log L$ disconnected subnetworks. This leads to a decrease in $\langle\delta v^2\rangle$ with increasing $L$ in the thermodynamic limit.
As shown in Fig.~\ref{fig:VelVar} $(a)$ and $(b)$, $\langle\delta v^2\rangle\times L$ captures the disconnected transition point around $\tilde M\approx 1.8/L$.
Remarkably, the LMG limit is also captured by $\langle\delta v^2\rangle\times L$ as a function of the rescaled connectance $(1-\tilde M)\times L$, as demonstrated in Fig.~\ref{fig:VelVar} $(d)$ and $(e)$.


\section{Spectral form factor}\label{sec:SFF}
In this section, quantum chaotic signatures are tested by the spectral form factor (SFF) analysis. In contrast to the local analysis, SFF probes both short and long-range spectral correlations. In this approach, one accesses the two-point correlation function via its Fourier transform. The SFF  is  a function of the continuous time parameter $t$,
    \begin{equation}\label{eq: SFF_def}
    \begin{split}
        \mathrm{SFF}(t)&=\frac{1}{N^2_\mathrm{av}}\left\langle\left\lvert \sum_{i=1}^{N_\mathrm{av}}e^{-i2\pi t E_i}\right\rvert^2\right\rangle_\mathrm{ER}\\
        &=\frac{1}{N^2_\mathrm{av}}\left\langle\left\lvert \sum_{i,j=1,\,i\neq j}^{N_\mathrm{av}}e^{-i2\pi t (E_i-E_j)}\right\rvert^2\right\rangle_\mathrm{ER}+\frac{1}{N_\mathrm{av}},
        \end{split}        
    \end{equation}
    where $N_\mathrm{av}$ is the number of the investigated eigenvalues. 
     The SFF provides a complementary characterization to measures of the local spectral statistics~\cite{SFFEarlyWilkie, SFFearlyChaosLeviandier, SFFEarlyAlhassid, Sierant_2020_SFFLevelStat, Gharibyan2018_RMT_QMChaos} and has been used to diagnose quantum chaos and MBL~\cite{ProzenVidamerSFF_MBLChaos, KulkraniMBLSFF2021}. It has further been investigated in systems with classically chaotic counterparts~\cite{Apollfo2024SFFQMChaosproposal, Adolfo_Multipartite_Chaos_2022PRR, ZollerMBQMChaosSFFPRX, ProzenKickedIsingQuasiSFF, ProzenPRXKickedTOPSFF} and in relation to the quantum decay of the survival probability~\cite{ Adolfo_Survival_Probability2017PRD, LeaSantosPwoerLawDecay2017,delCampo_SFFS_Survival2018, Santos_Thouless_Time_2019}.
    
      \begin{figure*}[t!]
    \includegraphics[width=1\linewidth,trim={.5cm 4cm 0 0},clip]{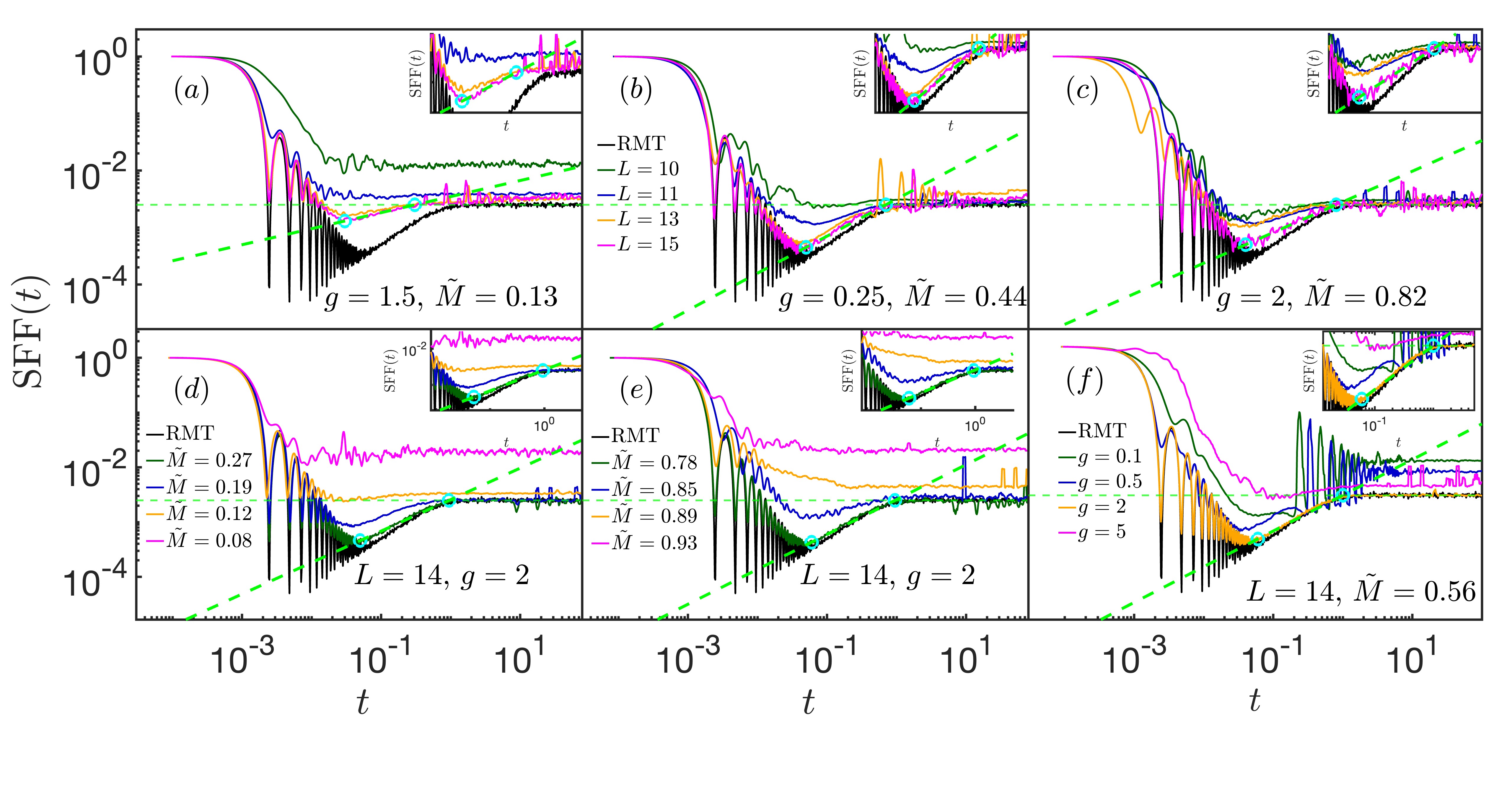}
    \caption{Spectral form factor for different $g,\,\tilde M$ and $L$ values for $N_\mathrm{av}=400$ energy levels around the middle of the spectrum. The linear ramp is indicated between $t_\mathrm{Th}$ and $t_H$ by a dashed line for comparison with the RMT result. Cyan circles show the positions of $t_H$ and $t_\mathrm{Th}$ while the horizontal dashed line highlights the long-time plateau of the SFF.
    Panels $(a)-(c)$ exhibit the breakdown of the chaotic SFF for different  $(\tilde M, g)$ pairs with varying $L$.    
    Panels $(d)$ and $(e)$ show similar breakdown for $L=14$ with varying $\tilde M$ for $g=2\,J$, whereby panel $(f)$ shows similar features varying $g$, for $\tilde M=0.56$. Insets show the close neighborhood of the correlation hole and the structure of the ramp.}\label{fig:SFF_examples}
\end{figure*}

\begin{figure*}[t!]
    \includegraphics[width = 1\linewidth,trim={6cm 2 52cm 0},clip]{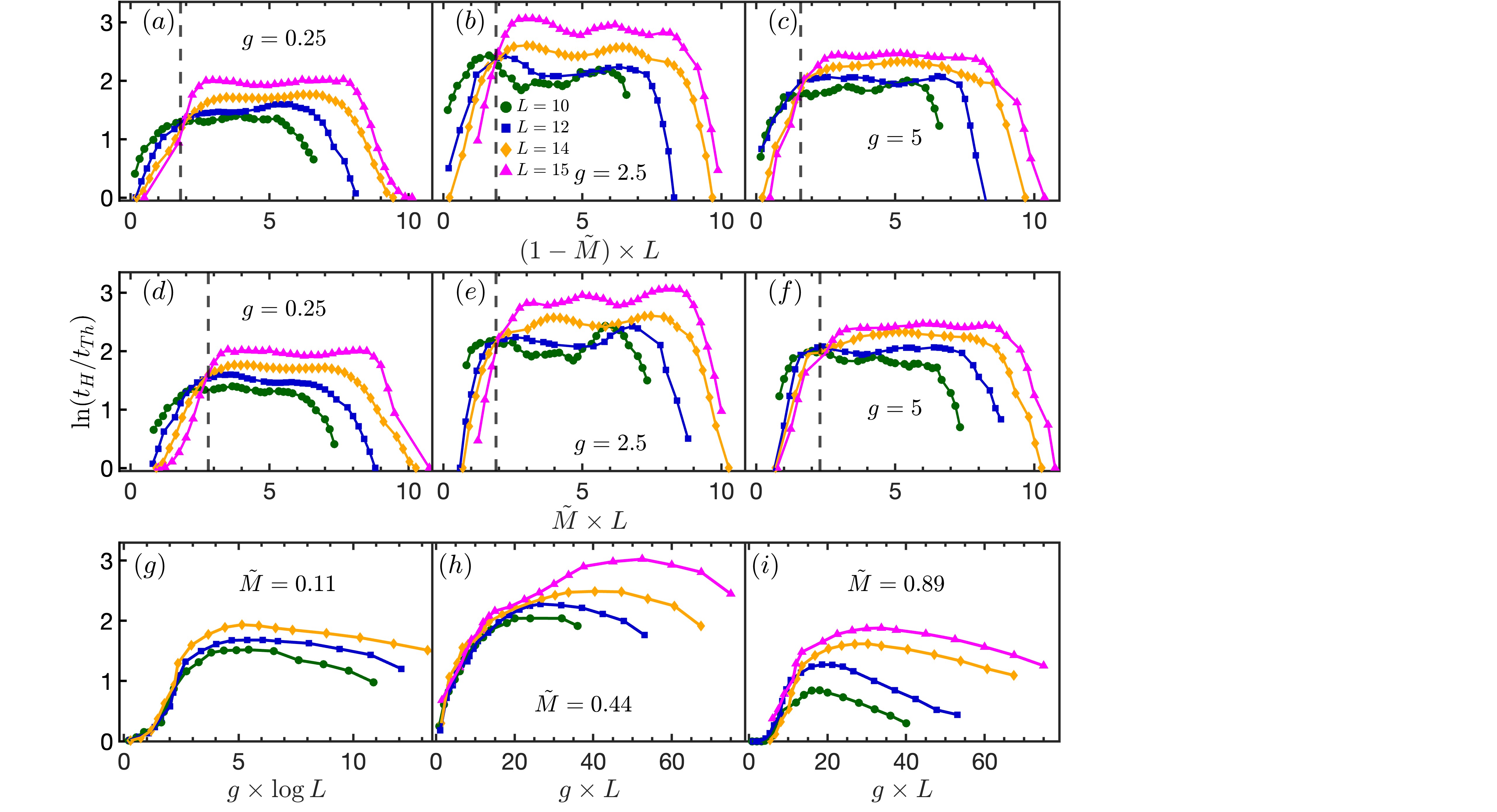}
    \caption{Logarithmic ratios of $t_H$and $t_\mathrm{Th}$ characterizing the correlation hole with varying $\tilde M$ and $g$ values for $L=10,\,12,\,14,\,15$.
    Panels $(a)-(c)$: Transition to the transverse field LMG limit as a function of $(1-\tilde M)\times L$ for $g=0.25\,J,\,2.5\,J,\,5\,J$.
    Panels $(d)-(f)$: Transition to the disconnected limit as a function of $\tilde M\times L$ for the same $g$ values. 
    The vertical dashed lines are indicated for the positions of the crossover towards the disconnected and LMG limits.
    Panels $(g)-(i)$: $g_H$ as a function of the rescaled $g$, for $\tilde M=0.11,\,0.44,\,0.89$ exhibiting the universal onset of quantum chaos. }\label{fig:g_H}
\end{figure*}

 For short time scales compared to the mean level spacing, $t\ll\delta\epsilon^{-1}$ spectral correlations at large separations dominate, resulting in a polynomial decaying shape. 
    In terms of the survival probability, initial short-time behavior corresponds to large energy scales, allowing for level-to-level transitions regardless of the spectral distance.
    This regime is governed by long-range level repulsion, inducing a characteristic oscillatory behavior for a chaotic spectrum.
    This ``dip" regime decays as $t^{-3}$ for chaotic and as $t^{-2}$ for the Poissonian regimes. 
    
    At intermediate time scales, the SFF exhibits a correlation hole, which is one of the most prominent signatures of quantum chaos governed by short-ranged level repulsion. The correlation hole sets the many-body Thouless time $t_\mathrm{Th}$~\cite{Santos_Thouless_Time_2019}. 
    The Thouless time, originally studied in the single-particle scenario, is identified with the inverse of the typical energy scale below which disordered quantum systems exhibit universal dynamics and strong similarities to random matrix theory. Recently, it has been generalized for many-body systems, in which it sets the time-scale needed to develop chaotic behavior~\cite{ProzenVidamerSFF_MBLChaos}.

    For $t>t_\mathrm{Th}$, quantum chaotic systems exhibit a linear ramp starting from the minimum of the correlation hole and saturating at $\mathrm{SFF}(t\rightarrow\infty)=\frac{1}{N_\mathrm{av}}$. This sets the relaxation or Heisenberg time $t_H$. It naturally captures the inverse of the mean level spacing $\delta\epsilon$ below which all level-to-level transitions are suppressed. Beyond $t_H$, the double sum of Eq.~\eqref{eq: SFF_def} converges to zero with increasing $t$ due to the cancellations of the randomly changing phases.
    In the Poissonian regime, the correlation hole does not emerge, and the short-time
    power-law decay immediately turns into a plateau. The absence of the correlation hole is equivalent to the breakdown of the quantum chaotic behavior.
    For disordered systems it also provides a prominent signature of the emerging MBL phase~\cite{ProzenVidamerSFF_MBLChaos, KulkraniMBLSFF2021, Herrera_Return_Prob_SFF}, while it is also a generic feature for integrable models and the same behavior is expected in the case of many-body scars~\cite{Turner2018_QM_MBScar, Serbyn2021_QM_MBScar, Chandran2023_QM_MB_Scars}.
    \begin{figure}[t!]    
    \includegraphics[width=1\linewidth,trim={0 10.5cm 0 0},clip]{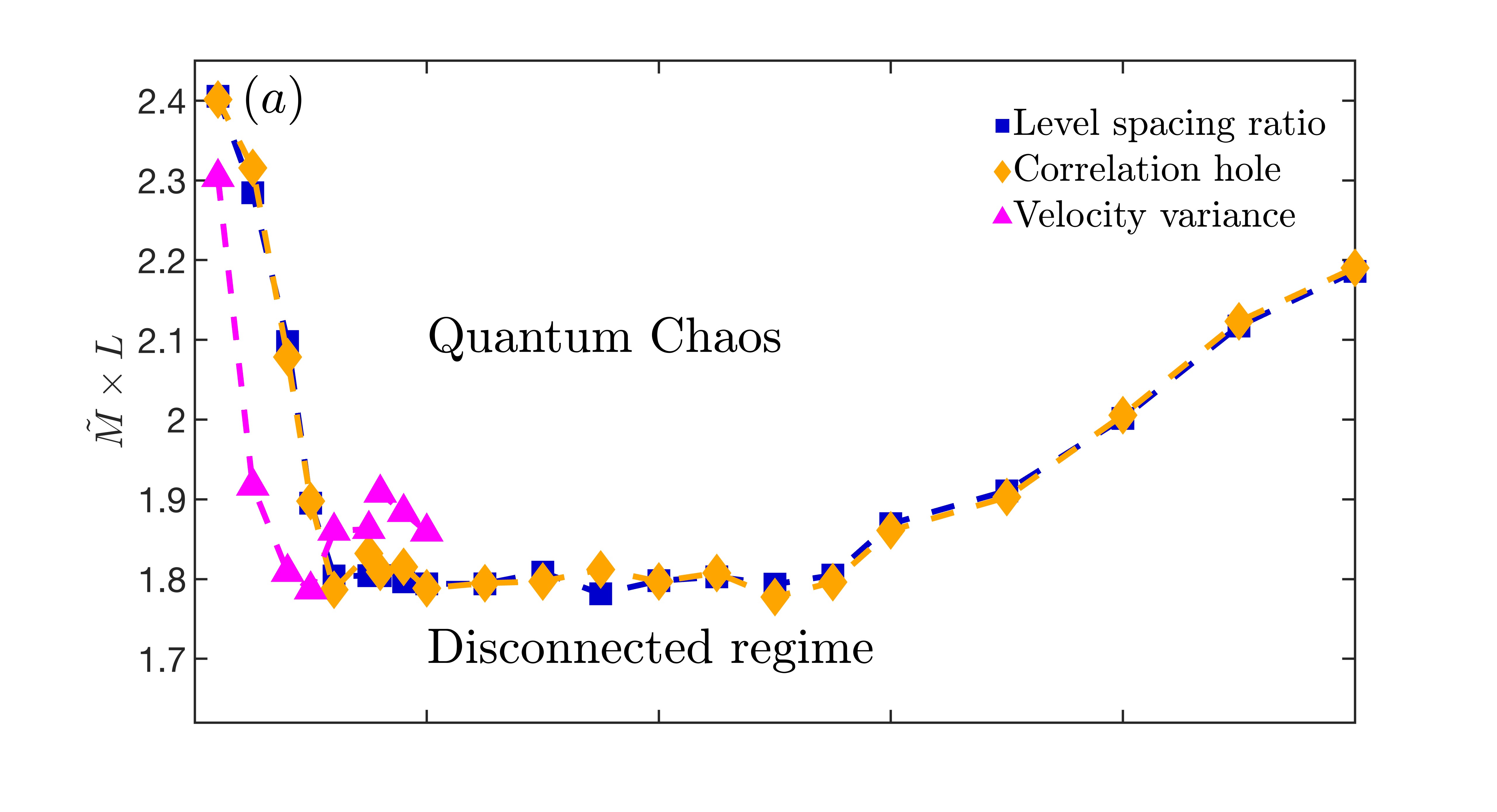}
    \includegraphics[width=1\linewidth]{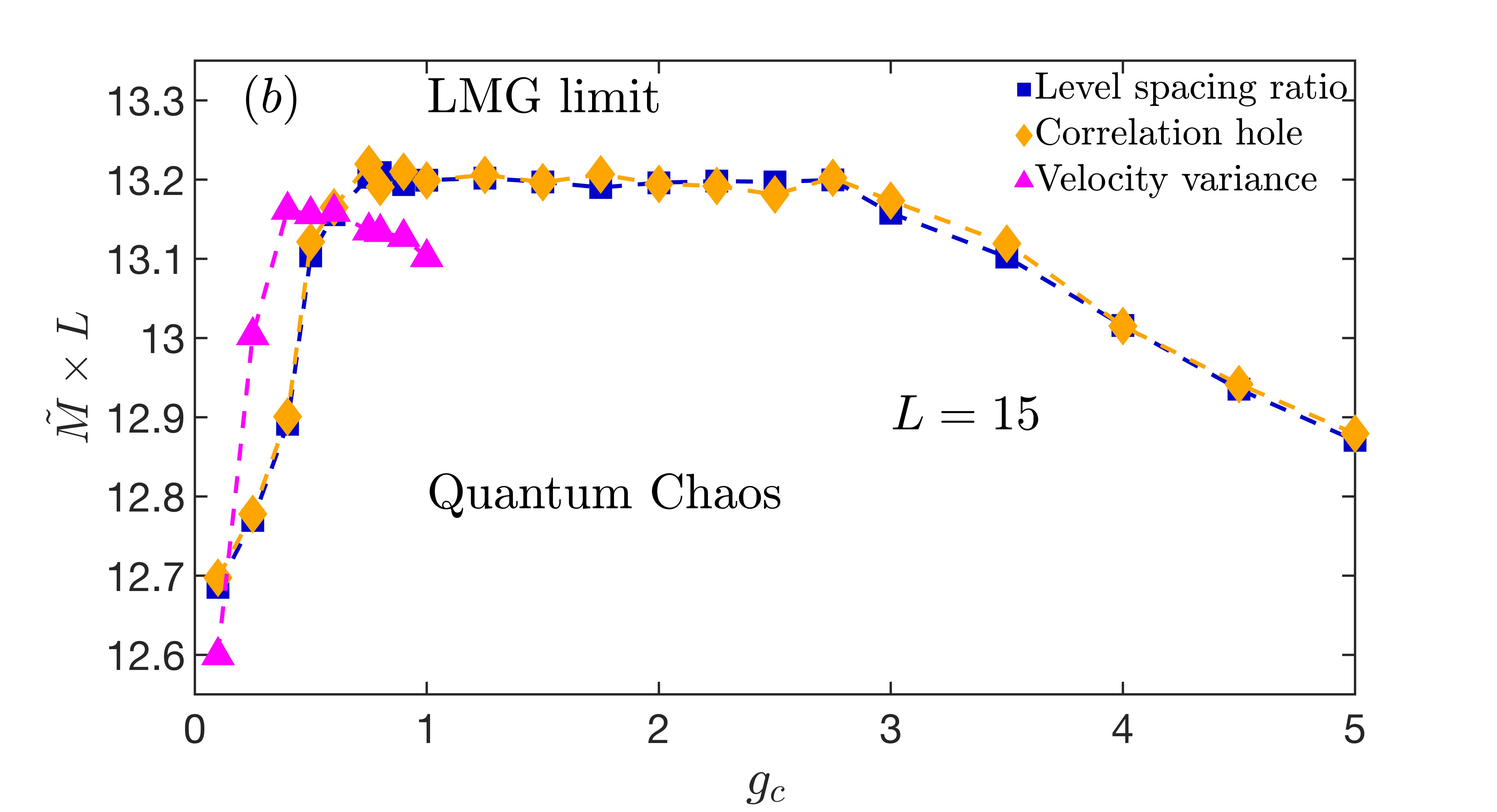}
    \caption{Phase diagram in the parameter space $(\tilde M,g)$, separating the quantum chaotic regions via the local and global measures of the level spacing ratio, the velocity variance, and the correlation hole. Panel $(a)$ shows the transition to the regime of disconnected networks as a function of the rescaled connectance $\tilde M\times L$.
    Panel $(b)$ shows the transition to the LMG integrable limit as a function of the rescaled connectance for $L=15$.
    }\label{fig:M_cg_c}
\end{figure}

The above picture changes for frustrated energy levels with exact or quasi-degeneracies. In particular, adjacent levels approaching much closer than $\delta\epsilon$ give rise to strong oscillations beyond $t_H$. Furthermore, exact degenerate pairs of eigenvalues with $E_i=E_j$ for $ i\neq j$ cancel out the  the phases, $e^{-i\,2\pi t(E_i-E_j)} = 1$. This leads to a constant increase of the plateau value scaling linearly with the number of degenerate eigenvalue pairs, $N_\mathrm{deg}$,
\begin{equation}
    \mathrm{SFF}(t\rightarrow\infty)=\frac{1}{N_\mathrm{av}}+2\frac{N_\mathrm{deg}}{N^2_\mathrm{av}}.
\end{equation}
While these effects break down the local spectral analysis, the correlation hole survives~\cite{Tezuka_BlackHoles_RMTSFF2017}. 
This allows for the identification of quantum chaos as demonstrated in Fig.~\ref{fig:SFF_examples} $(b), (e), (f)$.

The relative range of the correlation hole provides an efficient measure of the degree of chaoticity, captured by the logarithmic ratio of $t_H$ and $t_\mathrm{Th}$~\cite{ProzenVidamerSFF_MBLChaos} given by
    \begin{equation}
        g_H=\ln(t_H/t_\mathrm{Th})\,.
    \end{equation}
     It compares the relative size of the correlation hole to the corresponding RMT model with $N_\mathrm{av}$ investigated levels.

    For chaotic spectra, $g_H$ diverges with the number of levels, while for localized systems, it converges to zero in the thermodynamic limit.
    To focus on the width of the correlation hole, all SFF curves were normalized as $\mathrm{SFF}(t=0)=1$. The spectrum was unfolded to the energy range $[0,2^L\,J]$. This choice sets $t_H\approx 1/J$ around the middle of the spectrum. The minimum position of the correlation hole identifies $t_\mathrm{Th}$.
    
    To this end, quantum chaotic behavior was probed for the same data as via the local measures around the middle of the spectrum.
     As shown in Fig.~\ref{fig:SFF_examples} $(a)-(c)$, the short and the long-time behavior of the SFFs converges with increasing $L$ towards the corresponding RMT curve. 
     Similar to the spacing distribution, this convergence gets substantially weaker near the disconnected and fully connected regimes, Fig.~\ref{fig:SFF_examples} $(a),(c)$, as the correlation hole develops more slowly with the system size. 
     
     The dependence of the SFF curves on $\tilde M$ is also in good agreement with the predictions of the local analysis. 
     Increasing $L$ aids the system in escaping the limiting regimes of $g$ and $\tilde M$.
     The SFF exhibits a crossover to a Poissonian-like shape for low and high $\tilde M$. For low $\tilde M$, this happens on smaller time scales due to the smaller system sizes of the disconnected subnetworks, as demonstrated in Fig.~\ref{fig:SFF_examples} $(d)$ and $(e)$. In this regime, the degeneracies appearing in the peaks of the level spacing distribution around zero induce a constant shift of the plateau. 
     Similar to the level spacing ratios, $g_H\rightarrow0$ values capture the crossovers to the fully connected and the disconnected limits. These regimes are separated approximately at $\tilde M\approx\frac{1.8}{L}$ and $1-\tilde M\approx\frac{2}{L}$ for intermediate $g$. Moreover, these transition points get slightly shifted in the extreme transverse field regimes, as demonstrated in Fig.~\ref{fig:g_H} $(a)-(f)$. 
     The localization effects destroy the correlation hole for small and large $g$, as shown in Fig.~\ref{fig:SFF_examples} $(f)$. Additional oscillations appear beyond the ramp owing to the original degeneracies. In both limits, increasing $L$ restores the chaotic behavior and helps the system to develop a correlation hole. The logarithmic ratio $g_H$ follows approximately a universal curve as a function of $g\times\log L$ in the disconnected and $g\times L$  in the connected regimes. These findings are presented in Fig.~\ref{fig:g_H} $(g)-(i)$. Let us remark, that as in the case of the level spacing ratios, for extreme values of $g$ the correlation hole extends substantially slower than for intermediate values, as shown in Fig.~\ref{fig:g_H}  $(a),(c),(d)$ and $(f)$. Additionally, similar features are present for small and large values of $\tilde M$ exhibiting a considerably slower growth with $L$ compared to the intermediate connectance regime, Fig.~\ref{fig:g_H} $(g),(i)$.
     
The local and SFF measures allow us to spot the transition points in the space of $(\tilde M,g)$ as observed in Fig.~\ref{fig:M_cg_c}. For intermediate $g$, the scales $\tilde M\sim 2/L$ and $(1-\tilde M)\sim2/L$ separate the chaotic regimes from the disconnected and fully connected limits, respectively. These values become slightly larger when approaching the polarized and perturbative regimes. Here, $ \left\langle \delta v^2 \right\rangle$ remains a proper measure for $g\lesssim1$. Integrability is restored in the fully connected regime by reaching the transverse field LMG limit. Poissonian-like behavior naturally arises in the disconnected limit due to the independent subgraphs. However, the spectral nature of the small subnetworks preserves their chaotic behavior.

\section{Conclusions} \label{sec:Concl}
We have reported that random Ising networks exhibit quantum chaos, originating from the underlying Erd\H{o}s-R\'enyi graph topology. We investigated the level spacing distribution, the level velocity distributions, and the spectral form factor.

The spin Hamiltonian with ferromagnetic interactions and connectivity given by the Erd\H{o}s-R\'enyi random graph was mapped to a hopping model on an $L$ dimensional hypercube. The ratio of the on-site energy correlations and the averages allowed for the analytical treatment of the onset of quantum chaos as a function of the connectance and the transverse field. 
The level spacing distribution captures the transitions in the limiting regimes of the connectance.
In the disconnected regime, the Wigner-Dyson spectral statistics breaks down due to the independence of the disconnected parts. Increasing the connectance washes out randomness and develops integrable behavior as the fully connected limit is approached. However, these disconnected and LMG limits only survive for connectances converging to zero and unity, respectively, in the thermodynamic limit.
As for the transverse field dependence, a universal system size scale was computed by large-order perturbation theory and verified numerically. This approach captures the onset of quantum chaos via the condition of long-range delocalization.

Supplementing the level spacing analysis, the level velocity statistics was also employed as a local spectral diagnostics. In particular, Gaussian velocity statistics is found in the chaotic regime, which survives in the disconnected subnetworks as well. In contrast to the spacing distributions, level velocities are governed by short-range delocalization. Gaussian velocity statistics emerges when the instantaneous eigenstates have finite overlaps with $X$ basis states separated by Hamming distance of order $O(1)$. Thus level velocity captures different regimes of quantum chaos.
Remarkably, the velocity variance provided a new measure to capture the crossovers to both the disconnected and the fully connected regimes. 

Finally, we investigated level correlations via the spectral form factor at arbitrary spectral distances. The correlation hole provides an efficient quantum chaotic indicator in spectral regimes where level spacing distribution and velocity statistics exhibit slight deviations. The spectral form factor curves and the correlation hole exhibit similar conditions for the breakdown of quantum chaos. In particular, the correlation hole vanishes in the disconnected and the fully connected limits. The former originates from the independence of the subnetworks, not accounting for the remaining chaotic signatures.
The latter captures the LMG limit, similar to that observed in the spacing distribution and the velocity statistics. Furthermore, the correlation hole signals the same universal onset of quantum chaotic behavior when the transverse field is increased or decreased from the small and large limits, respectively.

This work has several natural extensions. The relation between the graph randomness and the efficiency of quantum annealing algorithms and methods of shortcuts to adiabaticity are interesting topics for future research. Exploring the velocity statistics and its variance in systems with an MBL phase also provides an exciting question. In addition, it is of interest to explore how dynamical signatures of the spectral statistics are modified in the presence of noise and coupled to a surrounding environment, which is ubiquitous in noisy intermediate scale quantum devices.

\acknowledgements
This research was funded by the Luxembourg National Research Fund (FNR), grant reference 17132054. F.J.G-R acknowledge financial support from Spanish MCIN with funding from European Union Next Generation EU (PRTRC17.I1) and Consejeria de Educacion from Junta de Castilla y Leon through QCAYLE project. For the purpose of open access, the authors have applied a Creative Commons Attribution 4.0 International (CC BY 4.0) license to any Author Accepted Manuscript version arising from this submission.


\onecolumngrid

\appendix

\section{On-site energy correlations}\label{App: Onsite_Energies}
In this appendix, we show the explicit calculations leading to the on-site energy correlations in Eq.~\eqref{eq:relative_correlators}, capturing the dependence on $L$ and $\tilde M$.
Let us consider two arbitrary spin configurations, such that the last $r$ spins appear with opposite signs, $\{\sigma_1,\dots,\sigma_L\},\,\{\sigma_1,\dots,\sigma_{L-r},-\sigma_{L-r+1},\dots,-\sigma_L\}$. For convenience, we use the variables $\sigma_i=\pm1$. This construction separates the two spin configurations by Hamming distance $r$. The corresponding on-site energies can be written as
\begin{equation}
\begin{split}
    &E=\sum_{i\neq j}A_{ij}\sigma_i\sigma_j=\sum\nolimits^{(0)}_{i,j}A_{ij}\sigma_i\sigma_j+2\sum\nolimits^{(r)}_{i,j}A_{ij}\sigma_i\sigma_j\,=E^{(0)}+E^{(r)},\\
    &\tilde E=\sum_{i\neq j}A_{ij}\sigma_i\sigma_j=\sum\nolimits^{(0)}_{i,j}A_{ij}\sigma_i\sigma_j-2\sum\nolimits^{(r)}_{i,j}A_{ij}\sigma_i\sigma_j=E^{(0)}-E^{(r)},\\
\end{split}
\end{equation}
where for brevity, we suppressed the $\frac{J}{M}$ normalization and the $z$ superindices. The last factor of $2$ is due to the symmetry of $A_{ij}=A_{ji}$. Here, the summations are understood as
\begin{align}\label{eq: truncated_sums}
    &\sum\nolimits^{(0)}_{i,j}A_{ij}\sigma_i\sigma_j=\sum_{i,j=1,i\neq j}^{L-r}A_{ij}\sigma_i\sigma_j+\sum_{i,j=L-r+1,i\neq j}^LA_{ij}\sigma_i\sigma_j\,,\\
    &\sum\nolimits^{(r)}_{i,j}A_{ij}\sigma_i\sigma_j=\sum_{i=1}^{L-r}\sigma_i\sum_{j=L-r+1}^LA_{ij}\sigma_j+\sum_{i=L-r+1}^L\sigma_i\sum_{j=1}^{L-r}A_{ij}\sigma_j\,.
\end{align}
In particular, the first accounts for both sums inside either the flipped or the identical sectors, while the second for the sums performed in different segments.

 Here, $E^{(0)}$ and $E^{(r)}$ correspond to the energies of the identical and the flipped segments of the spin configurations, respectively.
First, we compute the on-site average energies and variances,
\begin{equation}
\begin{split}
    &\langle E^{(0)}\rangle=\sum\nolimits^{(0)}_{i,j}\langle A_{ij}\rangle\sigma_i\sigma_j=p\sum\nolimits^{(0)}_{i,j}\sigma_i\sigma_j=p\left(\left(S^{(0)}\right)^2+\left(S^{(r)}\right)^2-L\right),\\
    &\langle E^{(r)}\rangle=2\sum\nolimits^{(r)}_{i,j}\langle A_{ij}\rangle\sigma_i\sigma_j=2p\sum\nolimits^{(r)}_{i,j}\sigma_i\sigma_j=2pS^{(0)}S^{(r)}\,.
    \end{split}
    \end{equation}
    Both expressions are the direct evaluations of the defining relations in Eq.~\eqref{eq: truncated_sums} as the adjacency matrix disappears from the summation due to the averaging.
    Here, $p$ denotes the probability of a link between two nodes, which converges to the connectance in the thermodynamic limit. In addition, $S^{(0)}=\sum^{(0)}_i\,\sigma_i,\,S^{(r)}=\sum^{(r)}_i\,\sigma_i$ denotes the total spins of the identical and flipped segments.
    Next, we consider the on-site variances, starting first with each segment separately,
    \begin{equation}\label{eq: flipped_and_identical_energies}
    \begin{split}
    \langle \left(E^{(0)}\right)^2\rangle&=\sum\nolimits^{(0)}_{i,j}\sum\nolimits^{(0)}_{k,l}\langle A_{ij}A_{kl}\rangle\sigma_i\sigma_j\sigma_k\sigma_l\\
    &=2p\sum\nolimits^{(0)}_{i,j}\sigma^2_i\sigma^2_j+p^2\left(\sum\nolimits^{(0)}_{i,j}\sigma_i\sigma_j-L\right)\left(\sum\nolimits^{(0)}_{k,l}\sigma_k\sigma_l-L\right)-2p^2\sum\nolimits^{(0)}_{i,j}\sigma^2_i\sigma^2_j,\\
    &=2p(1-p)((L-r)^2+r^2-L)+p^2\left(\left(S^{(0)}\right)^2+\left(S^{(r)}\right)^2-L\right)^2,\\
    \langle \left(E^{(r)}\right)^2 \rangle&=4\sum\nolimits^{(r)}_{i,j}\sum\nolimits^{(r)}_{k,l}\langle A_{ij}A_{kl}\rangle\sigma_i\sigma_j\sigma_k\sigma_l=8p\sum\nolimits^{(r)}_{i,j}\sigma^2_i\sigma^2_j+4p^2\sum\nolimits^{(r)}_{i,j}\sum\nolimits^{(r)}_{k,l}\sigma_i\sigma_j\sigma_k\sigma_l-8p^2\sum\nolimits^{(r)}_{i,j}\sigma^2_i\sigma^2_j,\\
    &=8p(1-p)r(L-r)+4p^2\left(S^{(0)}\right)^2 \left(S^{(r)}\right)^2,\\
    \langle E\tilde E \rangle&=\langle \left(E^{(0)}\right)^2 \rangle-\langle \left(E^{(r)}\right)^2 \rangle=2p(1-p)\left((L-r)^2+r^2-4r(L-r)-L\right)+p^2(S^2-L)(S^{(0)} -S^{(r)})^2-L),\\
    \langle \delta E\delta \tilde E\rangle&=\langle \left(E^{(0)}\right)^2\rangle-\langle \left(E^{(r)}\right)^2 \rangle-(\langle E^{(0)} \rangle^2-\langle E^{(r)} \rangle^2)=2p(1-p)\frac{(L-r)^2+r^2-4r(L-r)-L}{M^2}\\
    &\approx2p(1-p)\frac{L^2-6r(L-r)}{M^2}\sim L^{-2},
    \end{split}
\end{equation}
where in the last step, we restored the $1/M^2$ factor, and the total spin was denoted by $S=\sum_i\,\sigma_i=S^{(0)}+S^{(r)}$, which is by definition the sum of the total spins inside the identical and flipped sectors. Here, we also dropped the relatively vanishing $L$ factor in the numerator. The first term in $\langle \left(E^{(0)}\right)^2\rangle$ and $\langle \left(E^{(r)}\right)^2\rangle$ originates from the case where either $i=k,\,j=l$ or $i=l,\,j=k$. In this case, $A_{ij}$ and $A_{kl}$ belong to the same two vertices so their average is simply $\langle A_{ij}A_{kl}\rangle=\langle A^2_{ij}\rangle=p$ as $A_{ij}$ is either one or zero. If this is not the case, $A_{ij}$ and $A_{kl}$ behave as independent random variables, so $\langle A_{ij}A_{kl}\rangle=\langle A_{ij}\rangle\langle A_{kl}\rangle=p^2$. The second term stems from the case of no constraints between the pairs $(i,j)$ and $(k,l)$. In this latter, however, the summations are overcounting the cases of either $i=k,\,j=l$ or $i=l,\,j=k$. This is compensated by the third term.
As a result, the on-site energy variances read 
\begin{equation}
\begin{split}
    &\langle E^2\rangle=\sum_{i\neq j}\sum_{k\neq l}\langle A_{ij}A_{kl}\rangle\sigma_i\sigma_j\sigma_k\sigma_l=2p\sum_{i\neq j}\sigma^2_i\sigma^2_j+p^2\sum_{i,j}\sum_{k\neq l}\sigma_i\sigma_j\sigma_k\sigma_l-2p^2\sum_{i\neq j}\sigma^2_i\sigma^2_j=2p(1-p)L^2+p^2(S^2-L)^2,\\
    &\langle E\rangle=p\sum_{i\neq j}\sigma_i\sigma_j=p(S^2-L)\sim L^{-1},\\
    &\langle \delta E^2\rangle=\langle E^2\rangle-\langle E\rangle^2=\frac{2p(1-p)L^2}{M^2}\sim L^{-2},
    \end{split}
\end{equation}
following the same power law as the correlations and where the $1/M^2$ factor has been restored in the variance. Here, for the second moment of the total energy we followed the same steps as in Eq.~\eqref{eq: flipped_and_identical_energies}.

Finally, we investigate the averages and the variances of the energy differences between spin configurations separated by arbitrary $r$ Hamming distances:
\begin{align}\label{eq:App_d_Echange}
    &\langle E - \tilde E \rangle=4p S^{(0)}S^{(r)} \sim 4p\sqrt{r(L-r)},\\
    &\langle \delta(E - \tilde E)^2\rangle=\langle\delta  E^2\rangle+\langle\delta \tilde E^2\rangle-2\langle\delta E\delta \tilde E \rangle\\
    &\quad\quad\quad\quad\quad\quad=4p(1-p)L^2-4p(1-p)(L^2-6r(L-r)-L)\approx24p(1-p)r(L-r)\nonumber,
\end{align}
In the first equation, the sum and the difference of the identical and flipped segments of the two spin configurations simplify to the product of them. Around the middle of the spectrum, this typically scales as $S^{(0)}S^{(r)}\sim \sqrt{r(L-r)}$.
In the second equation, we dropped the single $L$ term again as it provides subleading corrections for $r\sim\sqrt L$ Hamming distances.
Note that both the averages and variances scale as $\sim\sqrt L$ for $r\sim\mathcal O(1)$ finite Hamming distances. However, for large $r\sim \sqrt L$ values, the variances  $\langle \delta(E-\tilde E)^2\rangle\sim L^{3/2}$ dominate over the averages $\langle E - \tilde E \rangle\sim L$. This feature will be central to our purposes in App.~\ref{App: Perturbative_Calc} .

\section{Perturbation theory approach of long-range delocalization}\label{App: Perturbative_Calc}
In this appendix, we use a large-order perturbation theory within the forward scattering approximation to verify the overall presence of chaotic spectral properties outside the limiting regimes $\tilde M$ and $g$. Chaotic spacing distribution is governed by correlated energy levels around the middle of the spectrum, which can be captured by the overlaps of computational basis states separated by Hamming distances $r\sim\sqrt L$.
For this, we consider the perturbation theory of order $r\sim \sqrt L$ to investigate the spread of the eigenstates among the most frequent spin configurations. 
Here, the dominant number of energy states accumulates within the range scaling as $\sim\sqrt M\sim L$ in the connected limit. This requires a minimum number of  $n\sim\sqrt L$ spin flips as the typical size of energy jumps scale as $\sim\sqrt L$. This is in agreement with the typical Hamming distance $r\sim\sqrt L$ between the corresponding spin configurations.

To get a lower bound on the overlap, we neglect spin-flip trajectories involving degenerate states, as they appear with a relative frequency suppressed as $\sim1/\sqrt L$. A spin flip leaves the energy invariant if its neighboring spins add up to zero. This can happen via $\binom{L}{L/2}\approx 2^L/\sqrt L$ ways out of all possible $2^L$ neighboring configurations, where the connectivity of the flipped spins has been replaced with their average of $M/L\sim L$ in the connected regime. Only the shortest spin trajectories are considered within the forward scattering approximation, which further validates neglecting spin flips between degenerate states. Thus, the non-degenerate perturbative correction reads
\begin{equation}\label{eq:T_n_app}
    T_n\sim\left(\frac{gM}{LJ}\right)^n\sum_{\mathcal P(\text{shortest paths})}\prod_{r=1}^n\frac{1}{\Delta_{\mathrm P_r}},
\end{equation}
with the sum running over the possible permutations of the shortest spin trajectories.
Here the scaling $(g/L)^n$ comes from the coefficient of the perturbation part, $\frac{g}{L}H_x$ and the $M$ factor originates from the normalization of $H_P$. The rescaled $\sim\mathcal O(J)$ energy differences after the $i^{\text{th}}$ steps are denoted by $\Delta_i$.

 Now, using the exact result of Eq.~\eqref{eq:App_d_Echange} and exploiting that the terms in the denominator in Eq.~\eqref{eq:T_n_app} on average increase apart from small fluctuations. This leads to $\Delta_r\sim\langle E_0-E_r\rangle\sim r^2-2rS\sim r\sqrt L$. Here, for finite $r$, the second term dominates, while for a large $r\sim\sqrt L$, the two terms become of the same order.
 Note that the variances are proportional to the average only in the long-range, $r\sim\sqrt L$ case, while for $r\sim\mathcal O(1)$ $\langle\delta\left(E_0-E_r\right)^2\rangle\sim\langle E_0-E_r\rangle^2$. This verifies further that the terms in the denominators of Eq.~\eqref{eq:T_n_app} can be characterized solely by the average in leading order.
 Thus, the typical scaling of the denominators read,
 \begin{equation}
     \prod_{r=1}^n\frac{1}{r\sqrt L\,J }\sim (n^2L)^{-n/2}\sim (JL)^{-n}\,.
 \end{equation}
 The numerator includes all possible jumps that increase the energy by $\sqrt L\,J$. The number of possible jumps in a typical scenario of reaching states at the opposite edges of the energy range $\sim L\,J$ scales down linearly with the already performed steps, leading to the scale $L(L-1)\dots(L-n+1)\sim L^n$.
Putting together the two competing terms arising from the number of possible spin trajectories and the suppressing energy denominators, we get,

\begin{equation}\label{eq:T_n_Z}
    T_n\sim\left(\frac{gM}{LJ}\right)^nL^n (n^2L)^{-n/2}\sim \left(\frac{gL}{J}\right)^n\,.
\end{equation}
Now, ergodicity is expected to break down when the long-range spreading becomes exponentially suppressed, i.e., perturbatively small in the thermodynamic limit, $g\lesssim L^{-1}J$. For $g\gtrsim L^{-1}J$, the perturbation expansion breaks down, signaling the onset of long-range delocalization. Thus, long-range exponential suppression is expected for smaller $g$ values than the breakdown scale of one spin-flip spreading. This indicates the absence of non-trivial localization and the dominance of quantum chaos whenever long-range suppression breaks down.

Next, let us consider the opposite limit when $g/L\gg J$ and the eigenstates are near the $X$ polarized spin states. Again, we consider the perturbation theory of order $n\sim\sqrt L$  matching the energy window of the dominant fraction of states around the middle of the spectrum. Here, single spin flips in the $X$ basis induce energy changes of $2g/L$.
Now, the $n^\text{th}$ order of perturbation expansion can be captured by random two spin-flip trajectories, neglecting the degenerate intermediate states.  As dictated by forward scattering approximation, these spin-flips provide energy-increasing contributions. In this case, the number of spin flips is restricted by a factor of $2$ but still leads to a numerator scaling as $M^n$. The denominator trivially scales as $\prod_{r=1}^n r\sim n^{n}$ leading to
\begin{equation}\label{eq:T_n_X}
    T_n\sim\left(\frac{LJ}{Mg}\right)^n \frac{M^{n}}{n^{n}}\sim 1\Rightarrow g\sim L^{1/2}J.
\end{equation}
This implies that at fixed $g$ values, increasing $L$ restores quantum chaos by helping the system escape the perturbative regime around the strongly polarized limit.
Both of the results in Eq.~\eqref{eq:T_n_Z} and Eq.~\eqref{eq:T_n_X} naturally carry on to the disconnected limit with $L$ replaced by $\log L$.

\section{Additional local spectral characteristics}\label{app:P_rStat}
\begin{figure*}[t!]    
    \includegraphics[width=\linewidth]{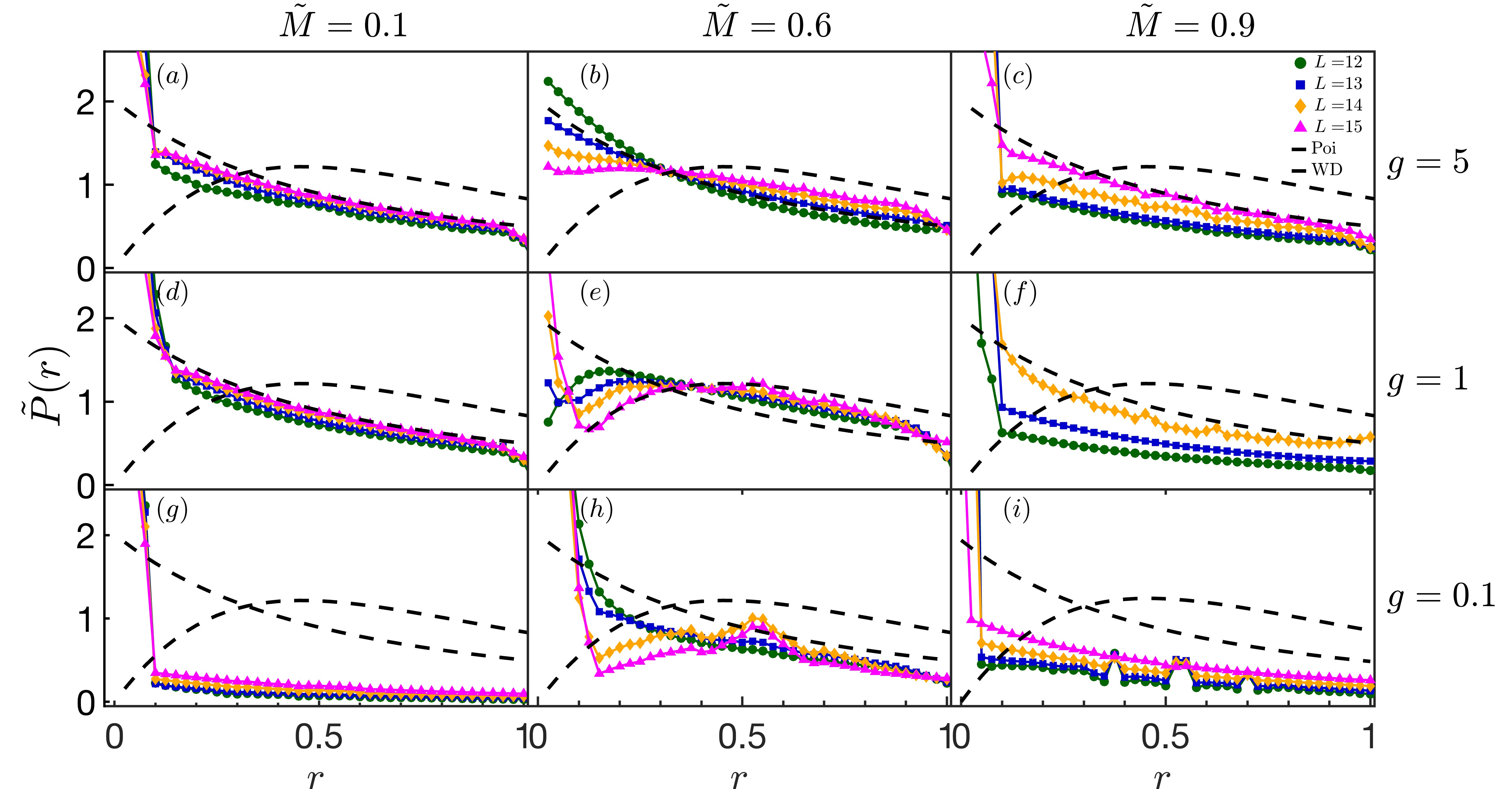} 
    \caption{Level spacing ratio distribution for $g=0.1\,J,\,J,\,5\,J$, $\tilde M=0.1,\,0.6,\,0.9$ and $L=12,\,13,\,14,\,15$. The energy window $[0.3\,J,\,0.4\,J]$ was considered in the unfolded spectrum.
    Panels $(a)-(c)$: $g=5\, J$ displaying the perturbative breakdown of quantum chaos, remedied by the increasing system size. Restoration of quantum chaos happens slower for low and high $\tilde M$ in $(a)$ and $(c)$, respectively.
    Panels $(d)-(f)$: Similar behavior for $g=J$ with a faster restoration of quantum chaos for intermediate $\tilde M$.
    Panels $(g)-(i)$: $g=0.1\, J$ exhibiting similar features with an overall slower convergence towards the WD distribution.\\
    }\label{fig: RStat_examples}
\end{figure*}

In this appendix, we provide further spectral characteristics of the level spacing distribution and level velocities.
We demonstrate the ubiquitous presence of quantum chaotic behavior by the level spacing ratio distribution. Expectedly, they exhibit the same qualitative behavior for the same $\tilde M$ and $g$ values. In short, the WD character breaks down at extreme $g$ limits, and the same features are also observed in the disconnected and densely connected regimes. The combination of these limits induces larger deviations from the chaotic behavior. However, the curves display an overall convergence towards the WD distribution with increasing $L$. As shown in Fig.~\ref{fig: RStat_examples} $(a),\,(d),\,(g)$, in the disconnected regime, the level spacing ratio distribution becomes more peaked around zero than the Poissonian one, similar to the spacing distribution. Here, the speed of the convergence towards the WD shape decreases substantially for large and small $g$ values. For intermediate $\tilde M$, the curves are much closer to the WD distribution. These features are shown in Fig.~\ref{fig: RStat_examples} $(b),\,(e),\,(h)$. Near the LMG limit, the chaotic character gets restored slower than for intermediate $\tilde M$ values; see Fig.~\ref{fig: RStat_examples} $(c),\,(f),\,(i)$.

\begin{figure*}[t!]
    \includegraphics[width=1\linewidth,trim={2.8cm 600 1.45cm 170},clip]{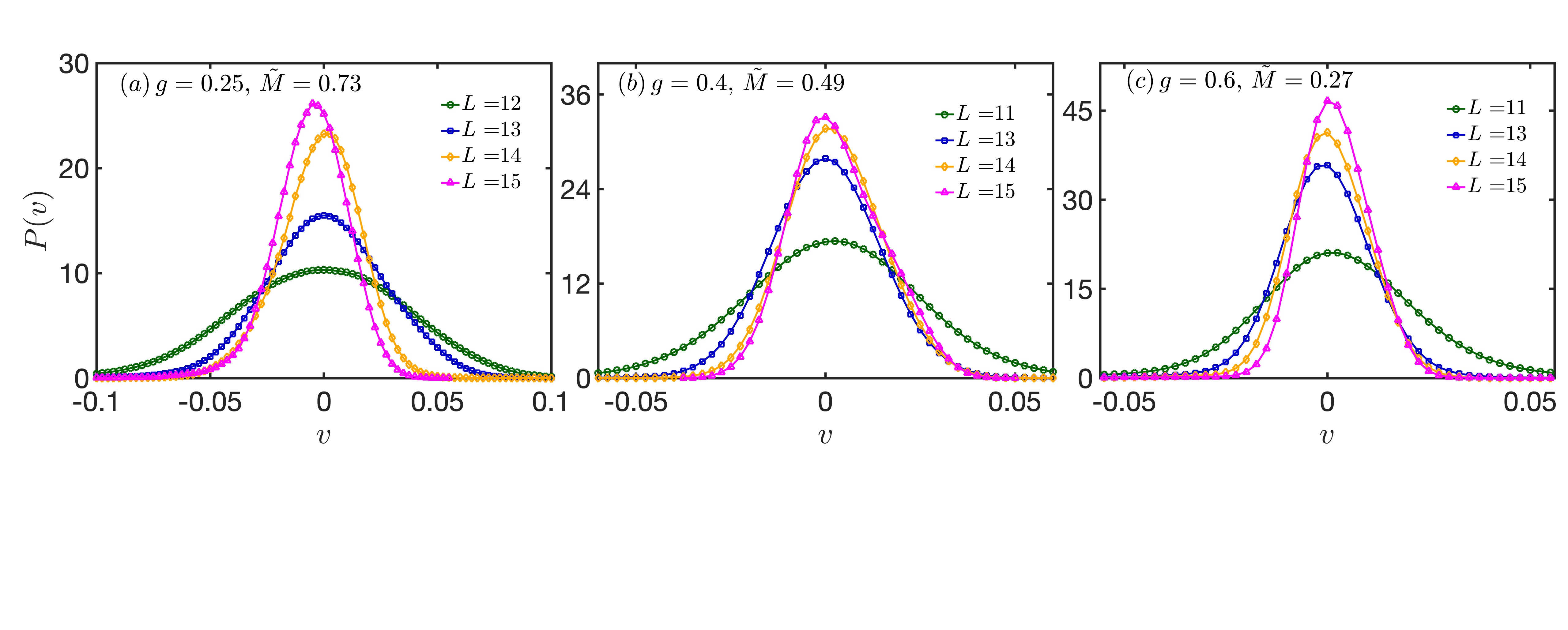} 
    \caption{Level velocity distributions for transverse field $g = 0.25J, 0.4J, 0.6J$ with connectance $\tilde M = 0.73, 0.49, 0.27$ in panels $(a) - (c)$ respectively. The system sizes ranging from small $L = 11$ to large $L = 15 $ are considered here. The distributions depicted here do not exhibit unit variance because these plots serve to illustrate the shape of Gaussian curves, which become narrower as the system size increases.} 
    \label{fig14}
\end{figure*}

As demonstrated in Fig.~\ref{fig14}, the velocity statistics follow Gaussian distributions, indicating chaotic signatures in the spectrum. Additionally, in agreement with Sec.~\ref{sec:VelStat}, the Gaussian curves get narrower, indicating a smaller variance with increasing $L$. However, extracting the precise size scale is beyond the numerical capacity. 

\section{Spectral independence of local measures and slight deviations in the upper half of the spectrum}\label{App: energy_spectrum}

\begin{figure*}[ht!]
    \includegraphics[scale=0.102,trim={205 215 1.1cm 20},clip]{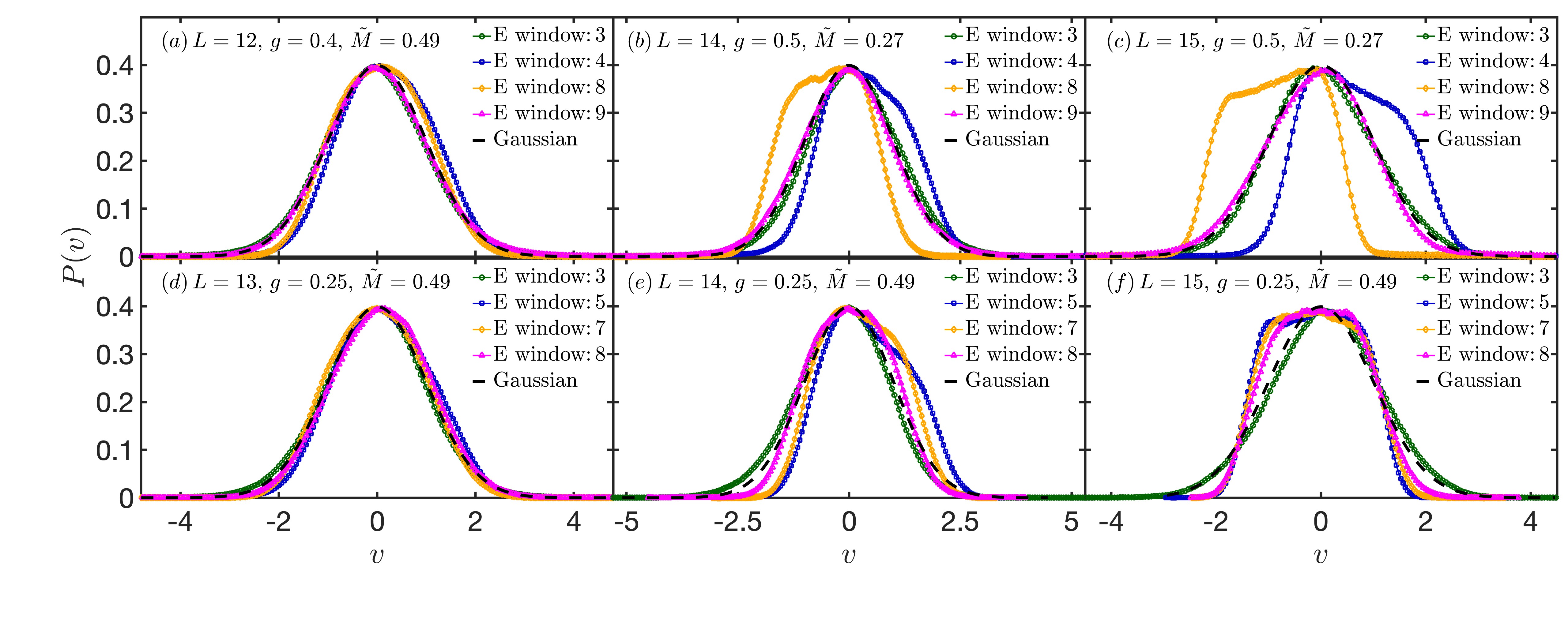}
    \caption{\label{fig:FIG8} Level velocity distributions for $\tilde M = 0.27,\,0.49$, $g = 0.25\,J\,, 0.4\,J\,, 0.5\,J\,$ and $L = 12,\, 13,\, 14,\,$ and $15$. The energy spectrum has been divided into $10$ energy windows. Panels $(a)-(c)$: $3^{\text{rd}},\,4^{\text{th}},\,8^{\text{th}},\,9^{\text{th}}$ energy windows.  Panels $(d)-(f)$: windows $3^{\text{rd}},\,5^{\text{th}},\,7^{\text{th}},\,8^{\text{th}}$. Frustration in the upper half of the spectrum induces slight deviations ($4^{\text{th}},\, 7^{\text{th}},\, 8^{\text{th}}$ energy windows).} 
    \label{1}
\end{figure*}

\begin{figure*}[ht!]
\includegraphics[width=1\linewidth]{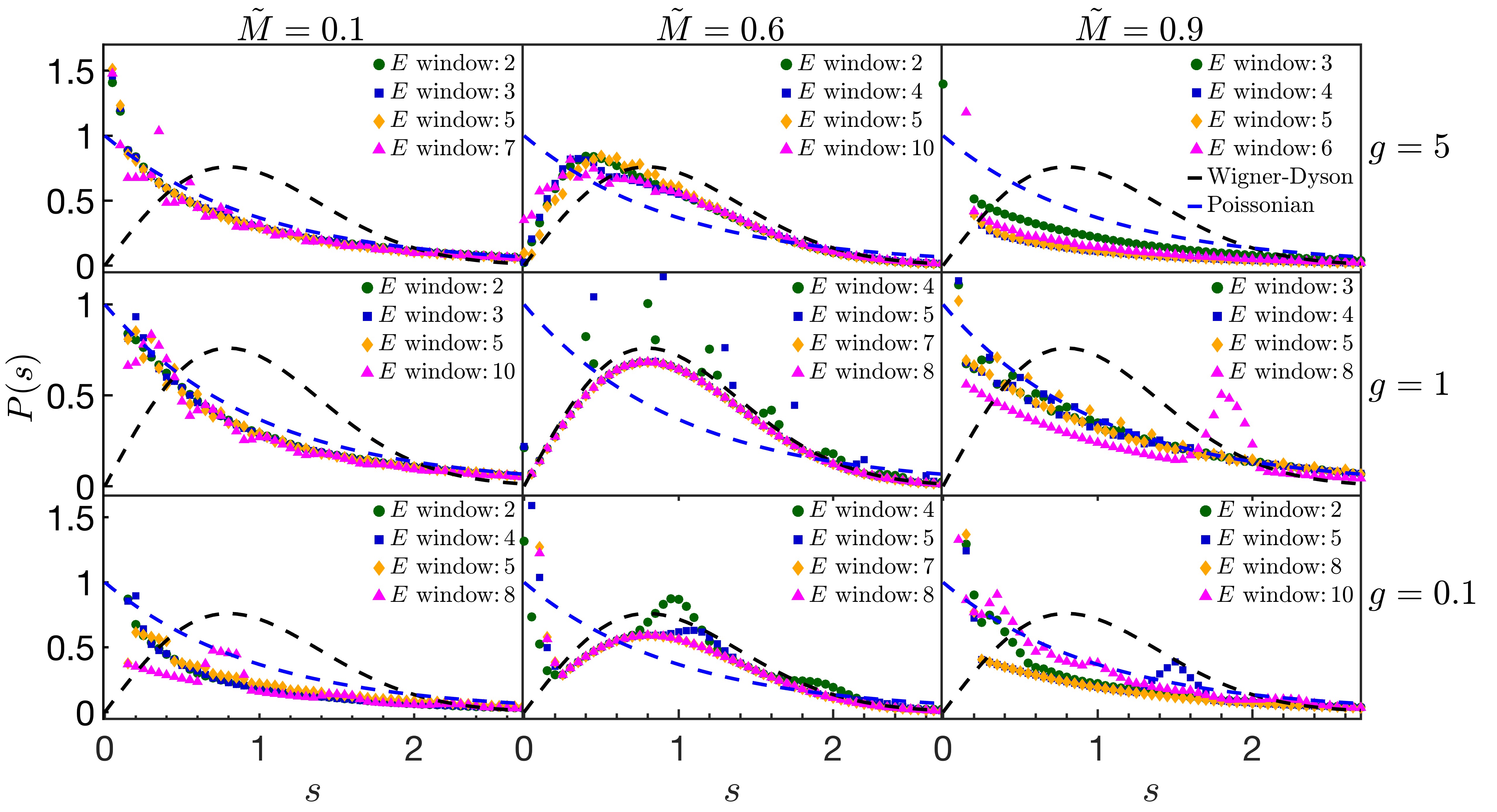}
\caption{Level spacing distribution in different spectral windows with $\tilde M=0.1,\,0.6,\,0.9$, $g=0.1\,J,\,J,\,5\,J$ and $L=15$. Slight deviations are present in the upper half of the spectrum.}
\label{fig13}
\end{figure*}

\noindent 
This appendix provides additional details about the spacing distribution and velocity statistics, investigated separately in ten energy windows in the unfolded spectrum.
As shown in Fig.~\ref{fig:FIG8} and Fig.~\ref{fig13}, the upper half around the middle of the spectrum exhibits a slight deviation from the traditional spectral characteristics, whereas the lower half remains consistent with the local measures. As illustrated in Fig.~\ref{fig13}, the level spacing distributions for $L = 15$ show noticeable deviations compared to the lower half of the spectrum. In Fig.~\ref{fig13}, spacing distribution in spectral regions $\tilde E\gtrsim J/2$ show additional peaks compared to the Poissonian and WD statistics, making it difficult to interpret the chaotic or integrable behavior of the spectrum. Moreover, the discrepancy from the chaotic behavior becomes more pronounced with increasing $L$ as shown in Fig.~\ref{fig:FIG8}. The velocity statistics for $L = 14, 15$ deviate from the Gaussian distribution, whereas for $L=12,13$ it follows approximately a Gaussian shape throughout the spectrum. Nevertheless, the deviations that appear in the upper half in Fig.~\ref{fig:FIG8} are of different kinds compared to Fig.~\ref{fig4}, Fig.~\ref{fig5}, and Fig.~\ref{fig6}. Here, the velocity distribution widens without any additional peaks, making it an efficient indicator of quantum chaos.

The observed deviations can be attributed to several factors. Frustrated loops can occur in connected graphs, leading to more degeneracies within the system. Additionally, certain graph configurations may involve nodes with even connectivity. Due to the uniform interaction strength, flipping these spins increases the number of degenerate excited states. These instances are more likely to occur around the middle of the spectrum at the highest density of states. The presence of degeneracies introduces a bias in the statistics, causing deviations from traditional spectral measures. In particular, the number of degenerate states is considerably lower for smaller $L$ compared to larger $L$. 


\section{Perturbative argument for the level velocity}\label{App: velocity_analytics}

\noindent The deviations of the velocity statistics from the Gaussian shape can be understood approximately using first-order perturbation theory. Here, we limit our analysis to the non-degenerate cases as the degenerate spin configurations separated by one spin flip emerge with a vanishingly small
frequency of $1 / \sqrt{L}$ in the thermodynamic limit, as discussed in App.~\ref{App: Perturbative_Calc}. For small $g$, $H_x$ acts as a perturbation to the unperturbed Hamiltonian $H_P$ in Eq.~\eqref{eq2}. 
Thus, the correction to the level velocity is given as,
\begin{equation}
    v^{(1)}_n = \frac{1}{L} \bra{n} {H}_x \ket{\delta n} = \frac{g M}{L^2}  \sum_{k \neq n} \frac{|\bra{n} {H}_x \ket{k}|^2} {\Delta E_{kn}} \label{eqE2}.
\end{equation}
Here, $\ket{k}$ and $\ket{n}$ are energy eigenstates of ${H}_P$, and $\Delta E_{kn}/M = E_k - E_n $ as the energy spectrum is rescaled to $1/M$ with $E_k$ and $E_n$ denoting the eigenenergies of $H_P$. 
The average energy difference between two spin configurations separated by a unit Hamming distance scales as $\Delta E_{kn}\sim S_{c}$. Here, the average connectivity of a single node scales as $c=M/L\sim {\tilde M L}$ in the connected limit. The total spin of the connected nodes, $S_{c}$ on average takes values within the range $[-\tilde ML/2, +\tilde ML/2]$ in agreement with the average connectivity. The most probable value of these spin configurations scales as $\sim \sqrt{\tilde ML}$. This is in agreement with Eq.~\eqref{eq:App_d_Echange}, where the fluctuations vary as $\sqrt{\tilde M L}$. Moreover, for each spin configuration $\ket{k}$, there exist $L$ possible one-spin-flip neighbors, $\ket{n}$, for which $\bra{n} {H}_x \ket{k}=1$. Hence, Eq.~\eqref{eqE2} becomes,
\begin{equation}
    v^{(1)}_n = \frac{g M}{L^2}  \sum_{k \neq n} \frac{|\bra{n} {H}_x \ket{k}|^2} {\Delta E_{kn}} \sim  \frac{g M}{L^2} \sum_{k = 1}^L \pm \frac{1}{\sqrt{\tilde M} L} \sim \frac{g M}{L^2} \frac{\sqrt{L}}{\sqrt{\tilde M}L}. \label{eqE4}
\end{equation}
The minus sign in the summation comes from the fact that the fluctuations dominate over the average. Due to the fluctuating sign but with the same order of magnitudes, the sum introduces a factor of $\sqrt{L}$. Using $\tilde M \sim M/L^2$ in the thermodynamic limit, Eq.~\eqref{eqE4} becomes,
\begin{equation}
    v_{n} \sim g \sqrt{\tilde M}. \label{eqE5}
\end{equation}
Now, let us investigate the opposite extreme, the fully polarized limit. In this regime, the level velocity in Eq.~\eqref{eq20} can be written as
\begin{equation}
    v_{n_x} = \frac{1}{L} \bra{{n_x}} {H}_x \ket{{n_x}} = \frac{E_{n_x}}{L} \label{eq28},
\end{equation}
where, $\ket{n_x}$ is an $X$ polarized eigenstate of ${H}_x$ with eigenenergy $E_{n_x}$. The spins in the paramagnetic phase can orient in $\pm X$ directions with $E_{n_x} \in [-L/2,+L/2]$. The degree of degeneracy is given by the binomial coefficients peaking at $E_{n_x} = 0$. This implies $E_{n_x} \sim \sqrt{L}$ in the middle of the spectrum for large $g$. Thus, Eq.~\eqref{eq28} becomes
\begin{equation}
    v_{n_x} \sim \frac{\sqrt{L}}{L} = \frac{1}{\sqrt{L}} \label{eq29} .
\end{equation}
Matching the two limits gives the result in the main text of Eq.~\eqref{eq26_1}.

\twocolumngrid
\bibliography{ref_v15}

\end{document}